\documentclass[twocolumn]{aastex6}
\usepackage{amsmath}

\makeatletter
\newcommand*{\rom}[1]{\expandafter\@slowromancap\romannumeral #1@}
\makeatother

\shorttitle{Data Reduction and Image Reconstruction Techniques for NRM}
\shortauthors{Sallum et al.}

\begin{document}


\title{Data Reduction and Image Reconstruction Techniques for Non-Redundant Masking}

\author{S. Sallum\altaffilmark{1}, J. Eisner\altaffilmark{1}}
\altaffiltext{1}{Astronomy Department, University of Arizona, 933
  N.\ Cherry Ave., Tucson, AZ 85721, USA}

\email{email: ssallum@email.arizona.edu}

\begin{abstract}
The technique of non-redundant masking (NRM) transforms a conventional telescope into an interferometric array.
In practice, this provides a much better constrained point spread function than a filled aperture and thus higher resolution than traditional imaging methods.
Here we describe an NRM data reduction pipeline.
We discuss strategies for NRM observations regarding dithering patterns and calibrator selection.
We describe relevant image calibrations and use example Large Binocular Telescope datasets to show their effects on the scatter in the Fourier measurements.
We also describe the various ways to calculate Fourier quantities, and discuss different calibration strategies.
We present the results of image reconstructions from simulated observations where we adjust prior images, weighting schemes, and error bar estimation.
We compare two imaging algorithms and discuss implications for reconstructing images from real observations.
Finally, we explore how the current state of the art compares to next generation Extremely Large Telescopes.

\end{abstract}

\section{Introduction}

Direct exoplanet studies rely on high contrast imaging methods used with adaptive optics systems. 
Techniques such as coronagraphy \citep[e.g.][]{2014ApJ...780..171G}, coupled with post-processing algorithms such as angular differential imaging \citep[ADI; e.g.][]{2006ApJ...641..556M} can detect planets around nearby stars \citep[e.g.][]{2015Sci...350...64M}.
The theoretical inner working angles of state of the art coronagraphs are $\sim 0.7-3 \lambda / D$ \citep[e.g.][]{2012SPIE.8442E..04M,2006ApJS..167...81G}.
In practice, the achievable inner working angle is not only a function of coronagraph design, but also of wavefront control.
Residual low order aberrations such as tip-tilt, which can be caused by seeing variations and vibrations \citep[e.g.][]{2010JOSAA..27A.122M}, can leak into the off-axis light \citep[e.g.][]{2012SPIE.8442E..04M}.
Furthermore, data reduction algorithms that build reference point-spread functions (PSFs) in the image plane \citep[e.g.][]{2007ApJ...660..770L,2012ApJ...755L..28S} perform poorly at small separations.
Close to the star, the small number of image elements cause small number statistics that degrade the achievable contrast \citep[e.g.][]{Mawet:2014}. 
These factors combine to limit achievable coronagraph performance to $\gtrsim \lambda / D$ \citep[e.g.][]{2012SPIE.8442E..04M,2017AJ....154...73R}.

Non-redundant masking \citep[NRM; e.g.][]{2000PASP..112..555T} is a way to probe smaller separations than more traditional imaging techniques such as coronography.
NRM turns a conventional telescope into an interferometer through the use of a pupil-plane mask.
No two baselines have the same position angle or separation, meaning that residual wavefront errors do not add incoherently and can be characterized.
Thus, despite blocking the majority of incident light, in the presence of noise NRM provides much better PSF characterization than filled-aperture observations.
It can detect sources with moderate contrast ($\sim 1:100$) at separations as small as $0.5~\lambda / D$, expanding the companion discovery phase space. 
NRM has led to the detection of stellar \citep[e.g.][]{2008ApJ...678L..59I,2012ApJ...753L..38B} companions, substellar companions \citep[e.g.][]{2012ApJ...745....5K,2015Natur.527..342S}, and circumstellar disk features \citep[e.g.][]{2015ApJ...801...85S,2015MNRAS.450L...1C} at or even within the diffraction limit.
This makes it particularly useful for direct planet formation studies, where most targets are at distances of $> 100$ pc.

Here we describe observational, data reduction, and image reconstruction techniques for NRM observations.
We discuss a \texttt{python} NRM data reduction pipeline, used primarily on observations from the Large Binocular Telescope Interferometer \citep[LBTI; e.g.][]{2008SPIE.7013E..28H} detector, LMIRCam \citep{2012SPIE.8446E..4FL}.
However our pipeline is general, and has also been applied to data from Magellan, Keck, and the VLT.

We show the effects of various calibration steps on the phases and squared visibilities.
We present reconstructed images from simulated observations of sources with different morphologies.
We discuss the effects of different initial images, error bar scalings, and baseline weighting schemes on these simulated reconstructed images.
We also compare simulated Giant Magellan Telescope reconstructed images to those currently achievable with the 23-meter LBTI.

\section{Experimental Setup}
In NRM observations, the detector records the interference fringes formed by the mask, called ``interferograms."
Fourier transforming the interferograms yields complex visibilities, which have the form $A \exp{i \phi}$.
Since the mask is non-redundant - no two hole pairs have the same separation and orientation - information from each baseline is located at a unique location in Fourier space. 
Sampling the Fourier transform, we calculate squared visibilities - the total power on each baseline - and closure phases - sums of phases around baselines forming a triangle \citep[e.g.][]{1958MNRAS.118..276J,1986Natur.320..595B}.
Closure phases are particularly powerful for companion detections since they are sensitive to asymmetries and are unaffected by atmospheric phase offsets.
Because closure phases are correlated, we then project them into linearly independent quantities called kernel phases \citep[e.g.][]{2010ApJ...724..464M,2013MNRAS.433.1718I,2015ApJ...801...85S}.
We apply both model fitting and image reconstruction to understand the source morphology.

\section{Observational Strategy}
NRM observations at LBT, Magellan, Keck, and the Very Large Telescope (VLT) are carried out in the near infrared ($\sim 1-4 ~\mu$m).
Since the sky background is high at these wavelengths (especially at wavelengths longer than 2 $\mu$m), we observe each target with two dither pairs per pointing.
We dither the images between the top and bottom halves of the detector.
To limit the effects of flat field errors, we keep the interferograms centered as close to the same pixels as possible for each dither position.
In clear conditions where the sky brightness is not variable on short timescales, one dither can be used to form the sky background for the other.

We alternate the science target observations with observations of unresolved calibrator stars.
This allows us to account for instrumental contributions to the closure / kernel phases and squared visibilities.
We match the target and calibrators' visible fluxes so that they have similar AO correction quality, and their infrared fluxes so they have similar noise levels on the science detector.
Calibrators are also vetted to ensure they do not have significant near-infrared excesses (indicating circumstellar dust) in their spectral energy distributions.
We choose multiple calibrators for each science target to minimize the possibility of contamination by binary calibrators.\\\\

\section{Data Reduction}

\subsection{Dark Subtraction, Flat Fielding, and Background Subtraction}
We create a master dark frame by taking the median of many dedicated dark images taken during the observing night.
We then dark subtract a set of sky flats taken during the night.
We median combine them and divide by the mode to create the master flat.

\begin{figure*}
\epsscale{0.75}
\plotone{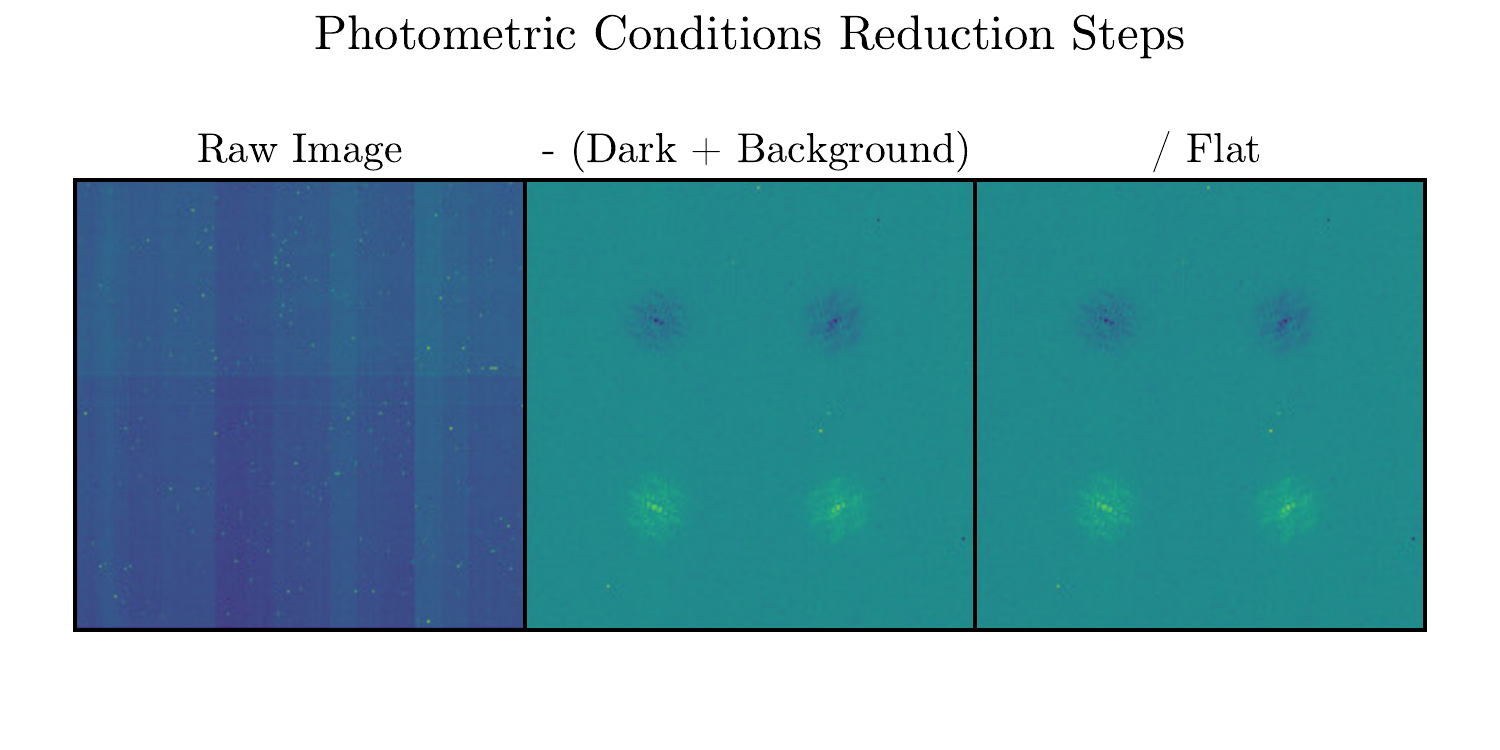}
\caption{Example raw science image (left), the same image after dark and background subtraction (center), and then after flat fielding (right). The left and center panels are shown with different color scales since the median value is significantly higher before background subtraction. The right panel is shown on the same color scale as the center panel. These images are from the December 2014 dataset published in \citet{2015Natur.527..342S}. In the two rightmost panels, the snowflake patterns show the interferograms formed by the mask. The two sides show the interferograms formed by the left and right LBT primary mirrors. The negative images on top show the results of using the median of one dither to subtract the sky background from the bottom dither.}
\label{fig-redp}
\end{figure*}

We apply dark, flat, and background calibrations differently depending on the observing conditions. 
In photometric conditions, for each dither pair we use one dither to perform dark and background subtraction on the other. 
We median all the frames in the first dither and subtract this median from each frame in the second dither. 
We then divide by a flat.
Figure \ref{fig-redp} shows example images as these steps are applied.

\begin{figure*}
\epsscale{1.0}
\plotone{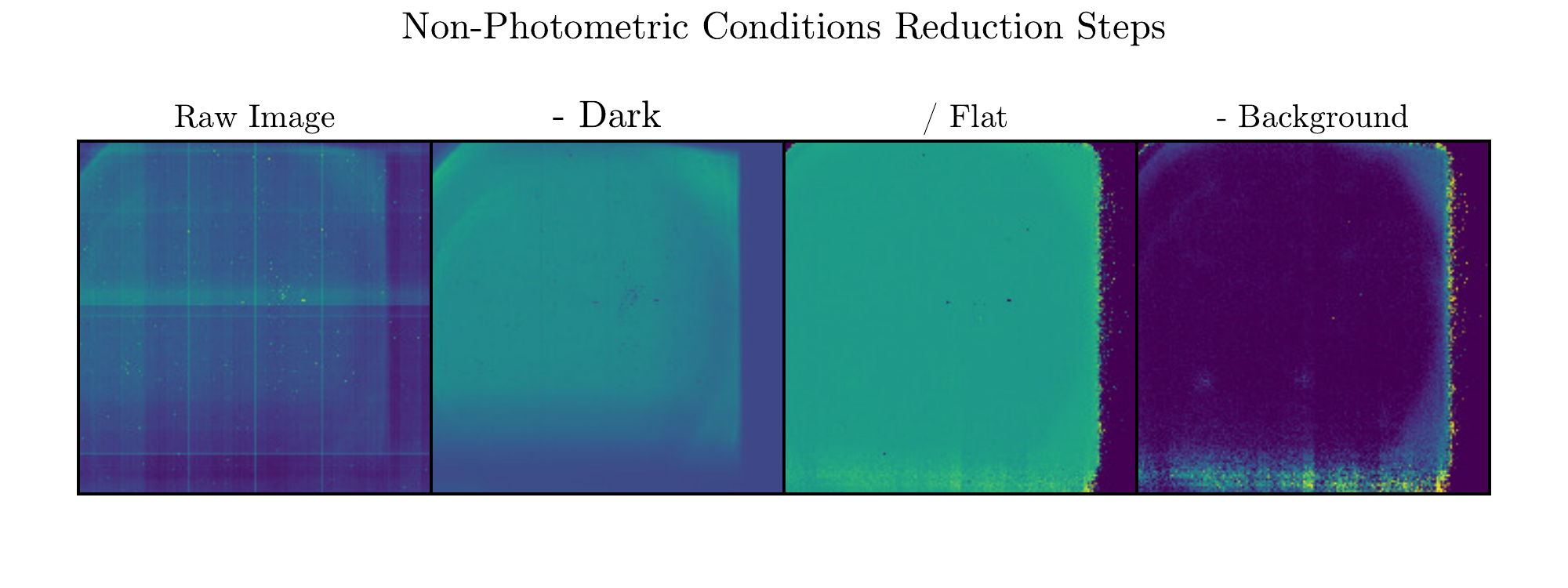}
\caption{From left to right: an example raw science image, the same image after dark subtraction, then after flattening, and lastly after background subtraction. Vignetting can be seen in the rightmost columns, bottom rows, and corners of the CCD. The center two panels are shown on the same color scales. The outermost two panels are on different color scales since the average pixel value changes significantly after dark and background subtraction. These images were taken after LMIRCam was upgraded to a 2048 $\times$ 2048 HAWAII-2RG detector.}
\label{fig-red}
\end{figure*}

For datasets taken in variable conditions such as intermittent cirrus, one dither may have a very different background level than another. 
In this case, rather than use one dither to perform the dark and background subtraction for another, we first subtract a median dark from every image in each dither.
We then divide by a flat.
After flattening we take the median of all pixels in non-vignetted portions of the CCD to be the median background signal. 
We subtract this median background value from the entire frame.
Figure \ref{fig-red} shows example images for each of these steps.

\subsection{Channel Bias Correction}

\begin{figure}
\epsscale{1.0}
\plotone{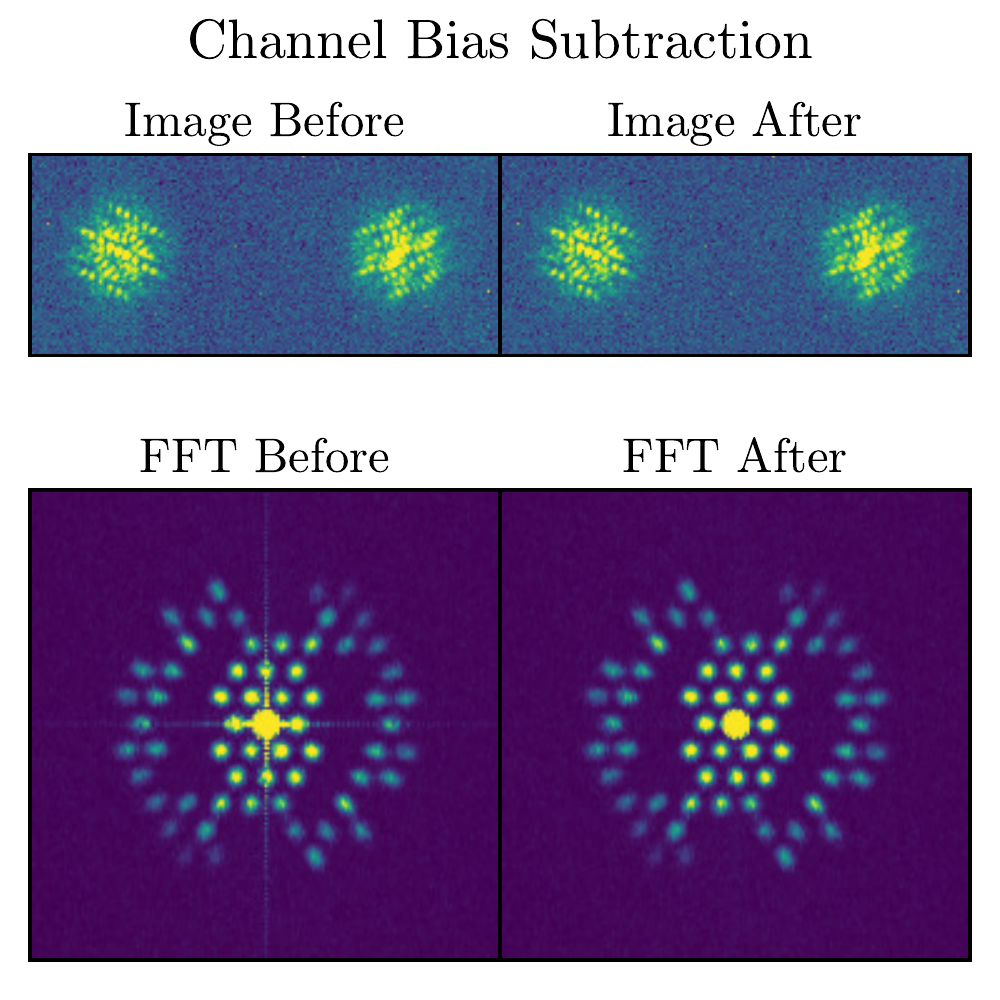}
\caption{Top: Images before and after channel bias correction. Bottom: Complex visibilities (Fourier transformed images) before and after channel bias correction. The vertical striping present in the uncorrected images creates low spatial frequency noise in the complex visibilities.}
\label{fig-channel}
\end{figure}

LMIRCam is a 2048 $\times$ 2048 HAWAII-2RG \cite[H2RG;][]{2008SPIE.7021E..0HB} detector, recently upgraded from a 1024 $\times$ 1024 HgCdTe detector \citep[e.g.][]{2012SPIE.8446E..4FL}.
Both of these configurations are read out in 64-pixel channels, each of which has its own analog-to-digital converter with a unique bias level. 
We correct for these different bias levels and any non-linear bias changes during each exposure by taking the median of each 64-column channel for each image.
We subtract the bias from each readout channel, as shown in Figure \ref{fig-channel}. 
If uncorrected, these channel biases lead to low spatial frequency noise in the complex visibilities (Figure \ref{fig-channel}).
While this 64-pixel scenario is specific to LMIRCam, H2RGs are used in other instruments with NRM capabilities such as the Gemini Planet Imager \citep[e.g.][]{2014SPIE.9147E..7OI}.
Regardless of the readout configuration, variable bias levels across the subframe may add systematic errors to the Fourier transform and can be corrected in a similar way.

\subsection{Bad Pixel Correction}

\begin{figure}
\epsscale{1.0}
\plotone{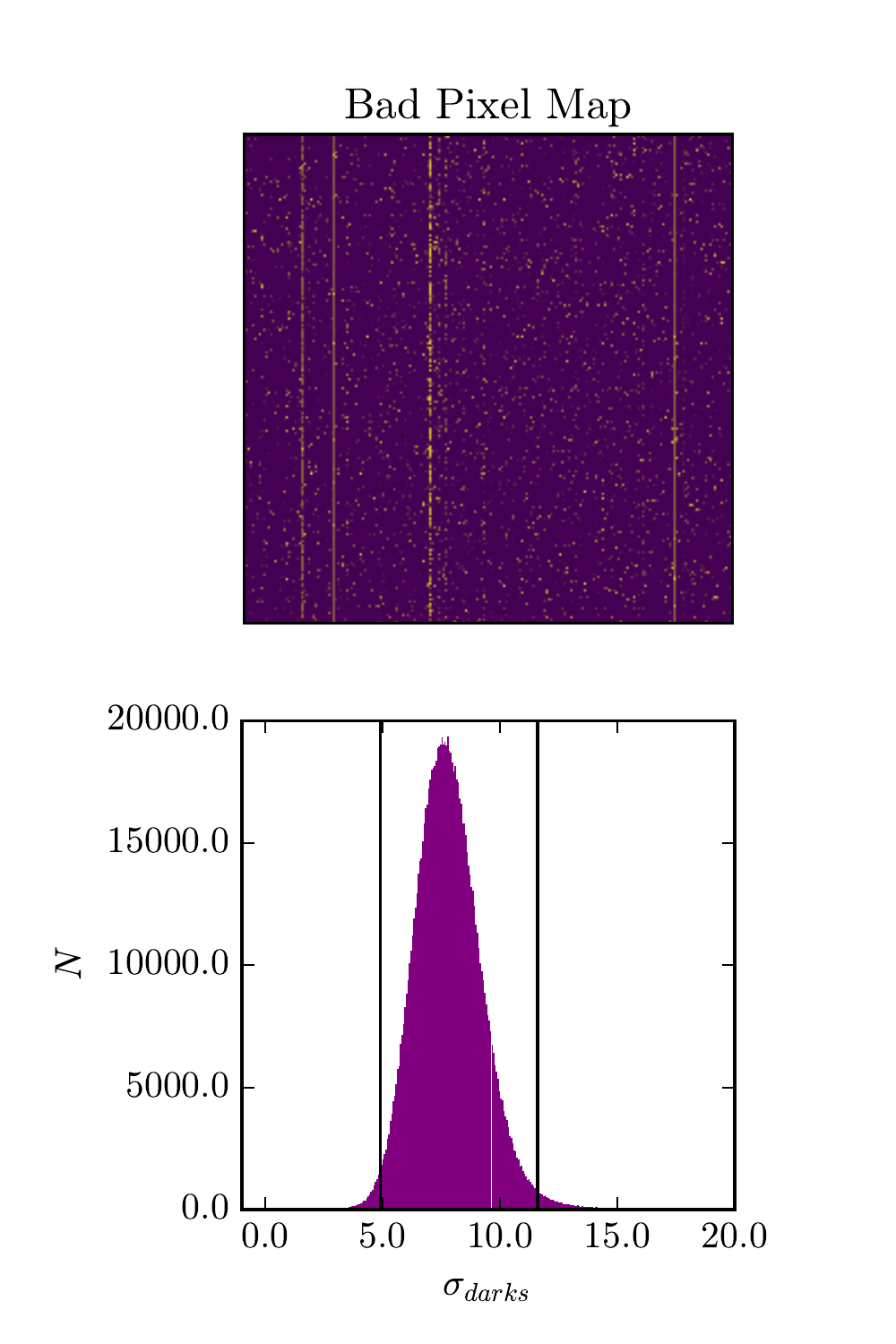}
\caption{Top: Bad pixel map created from a set of 20 dark frames. Bottom: Histogram of pixel standard deviations over 20 dark frames. The vertical lines indicate the cuts used to create the bad pixel map shown in the top panel. }
\label{fig-makebad}
\end{figure}

\begin{figure}
\epsscale{1.0}
\plotone{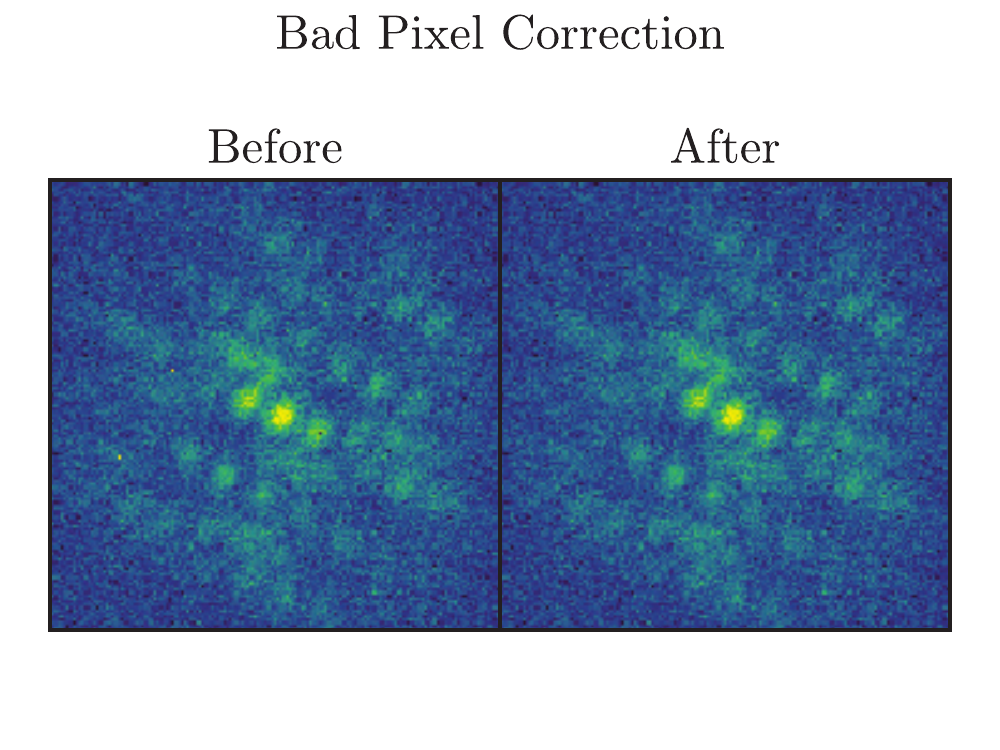}
\caption{Example interferogram subframe before (left) and after (right) bad pixel correction.}
\label{fig-badpix}
\end{figure}

We create a bad pixel map for each dataset using dark frames.
We calculate the standard deviation of each pixel across the cube of images.
We then label some fraction of pixels with the highest and lowest standard deviations as bad.
We allow for asymmetric cuts since the tails on the distribution of standard deviations may be asymmetric (they are for LMIRCam, see Figure \ref{fig-makebad}).
We use a variety of upper and lower cuts, and choose the bad pixel map that minimizes the scatter in the calibrator observations for the night.
We make corrections in the image plane, replacing each bad pixel with the average of the surrounding ones that have not been flagged.
For an example dataset taken in December 2014 \citep[published in][see Figure \ref{fig-badpix}]{2015Natur.527..342S}, dropping the top 2\% and bottom 1\% of pixels resulted in the best bad pixel correction.\\

\subsection{Noise Versus Reduction Steps}

\begin{figure}
\epsscale{1.0}
\plotone{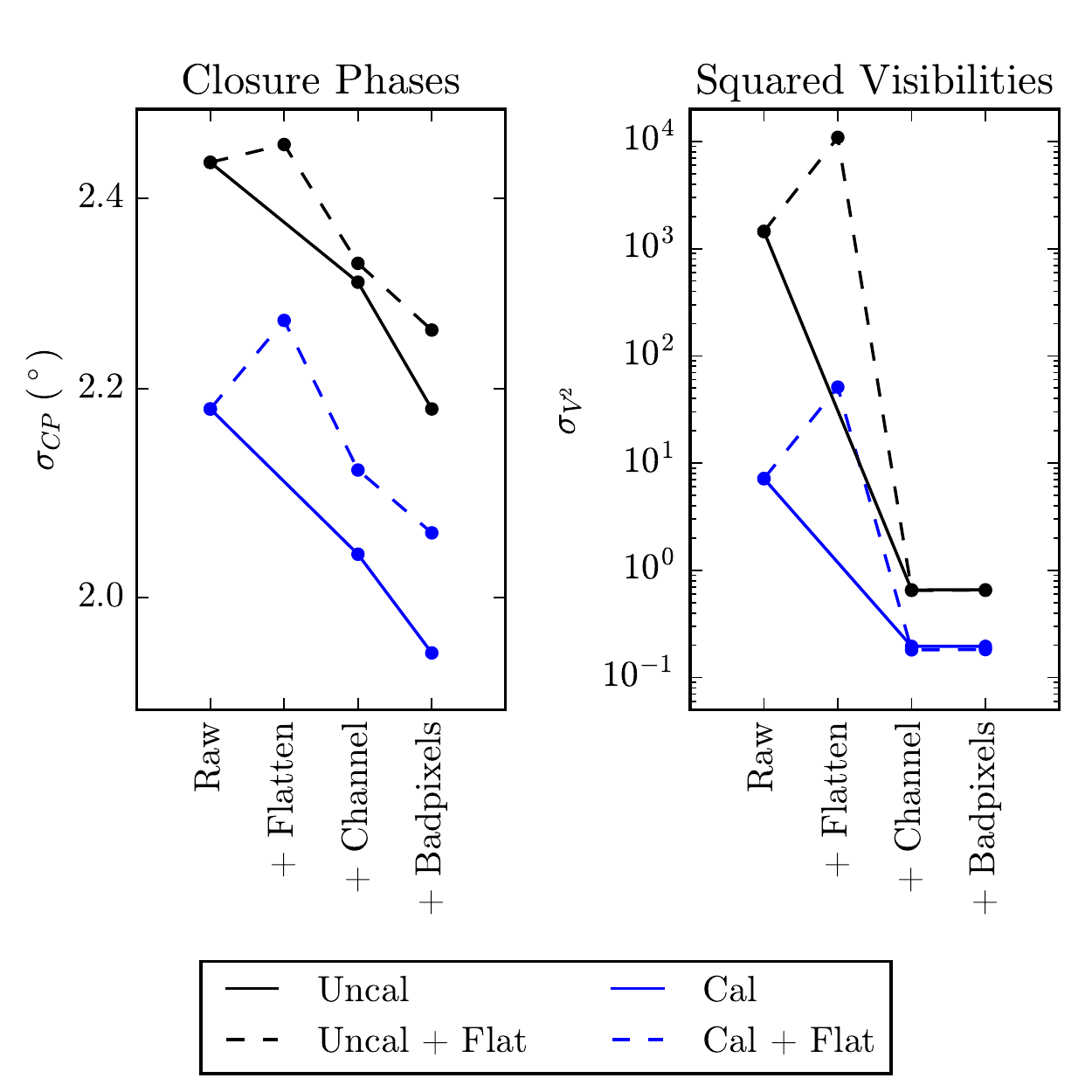}
\caption{Comparison of closure phase (left) and squared visibility (right) scatters for observations of unresolved stars from December 2014. Dashed lines show reductions that include flattening, while solid lines show reductions without flattening. Calibrated data are shown in blue, while uncalibrated data are shown in black.}
\label{fig-redcomp}
\end{figure}

In Figure \ref{fig-redcomp} we compare the scatter in the calibrated and uncalibrated data at each reduction step for the December 2014 LBT observations published in \citet{2015Natur.527..342S}.
We use two unresolved stars to calibrate each other, and then examine the scatter after flattening, channel bias correction, and bad pixel correction.
For the closure phases, the channel bias and bad pixel corrections decrease the closure phase scatter by nearly equal amounts ($\sim0.11^\circ-0.14^\circ$).
For the squared visibilities, reductions with a channel bias correction have orders of magnitude more scatter than those without a channel bias correction.
This is because only certain baselines sample the spatial frequencies of the channel bias noise; these will have much more power in them than the baselines that do not.

For the December 2014 L$'$ data, both the uncalibrated and calibrated data have lower scatter when a flat is not applied. 
We suspect this is because of the small number of flats used to create the master sky flat, which can be thought of as an image of ones with Gaussian noise added.
Flattening will thus add noise to the uncalibrated complex visibilities.
If the image on the detector always falls on the exact same pixels, then the flat field applied to the subframed interferogram is always the same.
With pointing inconsistencies, the subframed flat field is also inconsistent; any noise added by the master flat will then change slightly between pointings and will be more difficult to calibrate out. 
This increases the scatter in the calibrated data as well.
A large number of flats would decrease the noise in the master flat, and thus in both the uncalibrated and calibrated observables.
Previous simulations of NRM observations suggest that $\sim10^6$ photoelectrons per pixel are required for flattening to add less than $\sim0.06^\circ$ \citep{2013MNRAS.433.1718I}.
While flattening does not significantly change or improve the LMIRCam closure phases and squared visibilities, it may be more useful on other detectors with greater flat field variations across the subframed interferograms.

\section{Closure Phase Calculation}

The ($u,v$) coordinates for a closing triangle of baselines satisfy the following:
\begin{equation}
\left(u_1,v_1\right) + \left(u_2,v_2\right) + \left(u_3,v_3\right) = 0.
\label{eq-cp}
\end{equation}
To calculate a closure phase for a single triangle, we sample the complex visibilities for the three baselines and multiply them to form the bispectrum.
For each triangle, we average the bispectra measured from all images in a dither and then take the average bispectrum phase as the closure phase.
Since the bispectrum has both amplitude and phase, averaging bispectra upweights higher signal-to-noise measurements, which have higher complex visibility amplitudes.

The simplest way of calculating a closure phase in practice is to use only a single pixel for each baseline.
However, the finite mask hole size causes information from each baseline to be spread over multiple pixels in the Fourier transform.
To use information from more than just the central pixels, we can multiply the subframed interferogram by a window function.
Taking the Fourier transform of a windowed image convolves the Fourier transforms of the interferogram and window function, creating inter-pixel correlations.
We can then still sample single pixels but incorporate more information than the unwindowed single pixel method.
One caveat with this method is that narrow enough window functions can cause signals from adjacent baselines to bleed into one another.

\begin{figure}
\epsscale{1.0}
\plotone{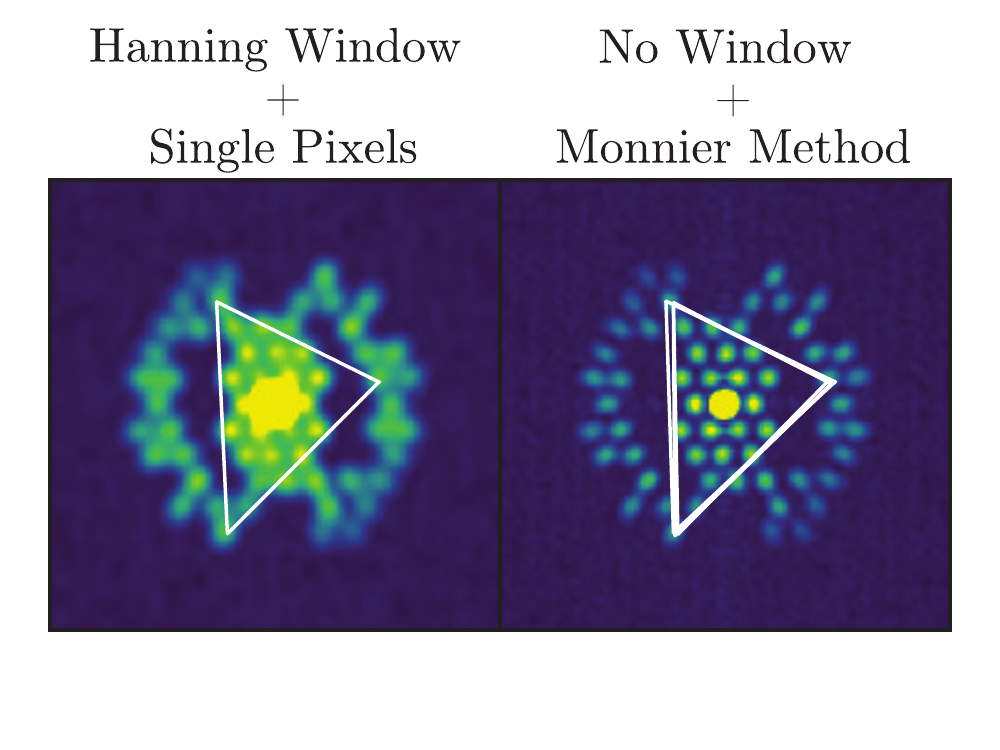}
\caption{Closure phase generation methods. The left panel shows the Fourier transform of an interferogram multiplied by a Hanning window. Here we would use a single pixel at the center of each splodge to calculate the bispectrum for a mask hole triplet (white lines). The window function shown here caused correlations between adjacent splodges. The right panel shows the Fourier transform of an unwindowed interferogram. The white lines show example closing triangles for individual pixels within each splodge. Here we average the bispectra of these triangles to calculate the bispectrum for each mask hole triplet.}
\label{fig-hannVsMonn}
\end{figure}

An alternative to windowing is to average many bispectra in each individual image for each triangle of baselines \citep[the ``Monnier" method; ][]{1999PhDT........19M}, as diagrammed in Figure \ref{fig-hannVsMonn}. 
We first take the Fourier transform of the unwindowed interferogram.
For each mask hole triplet we find all closing triangles that connect the extended signals from the three baselines.
We calculate their bispectra and average them to create a single bispectrum for each triangle of baselines in an image.
As in the single pixel method, we average the bispectra for all images in a dither and then take the bispectrum argument to be the average closure phase. 

\begin{figure}
\epsscale{1.0}
\plotone{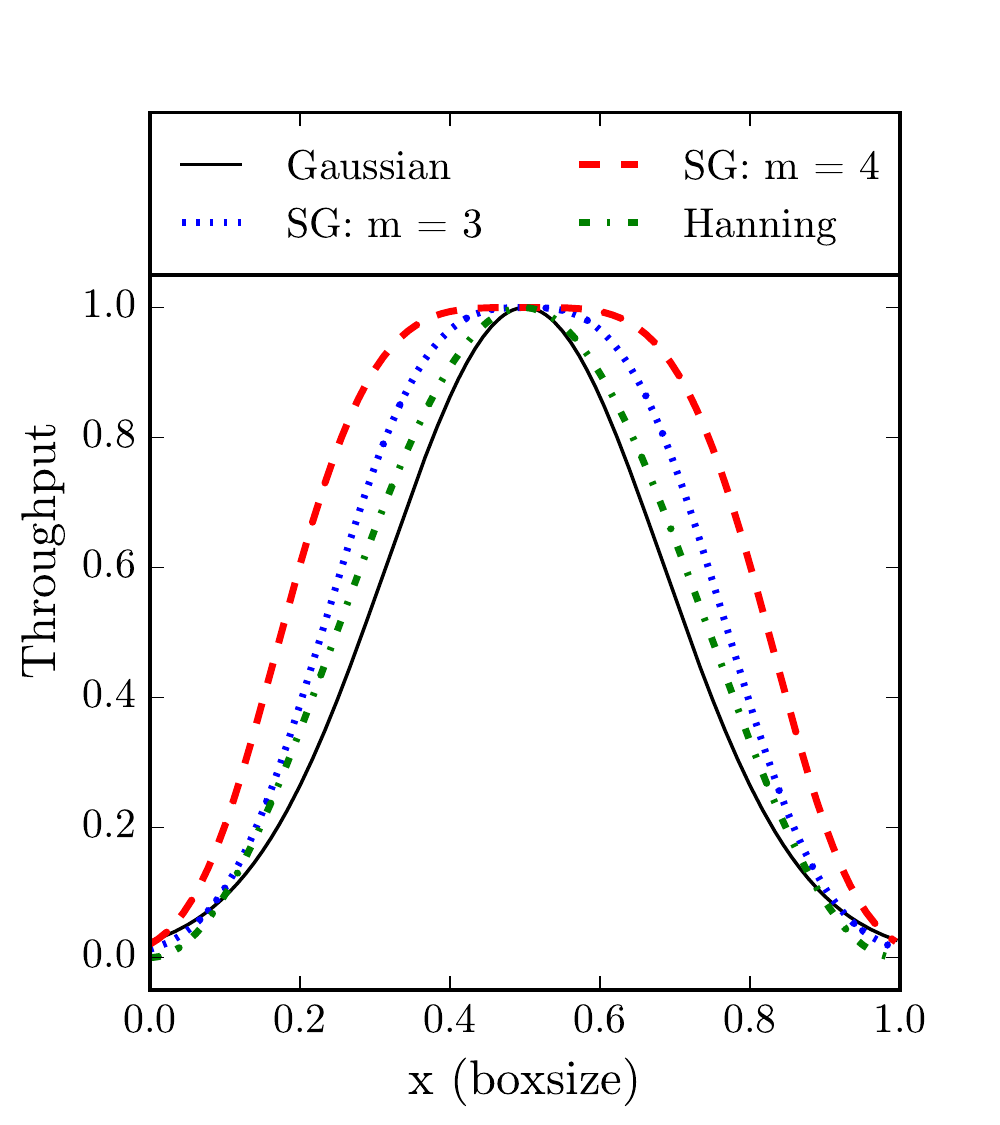}
\caption{Example window functions.}
\label{fig-windowcomp}
\end{figure}

Figure \ref{fig-windowcomp} shows example window functions (Hanning and super-Gaussian), and Figure \ref{fig-hannVsMonn} illustrates the different closure phase calculation methods.
A one-dimensional Hanning window of size M has the form
\begin{equation}
0.5 - 0.5 \cos{\frac{2 \pi n}{M-1}}
\end{equation}
where n goes from 0 to M-1.
To create a two-dimensional window, we make two one-dimensional Hanning windows and then take their outer product. 
We generate super-Gaussian windows according to the following:
\begin{equation}
\exp \left[-\ln\left(0.5\right)\left(\frac{r}{HWHM}\right)^m\right],
\end{equation}
where $r$ is the distance from the center of the subframe, and $HWHM$ the window half width at half maximum.

\begin{figure}
\epsscale{1.0}
\plotone{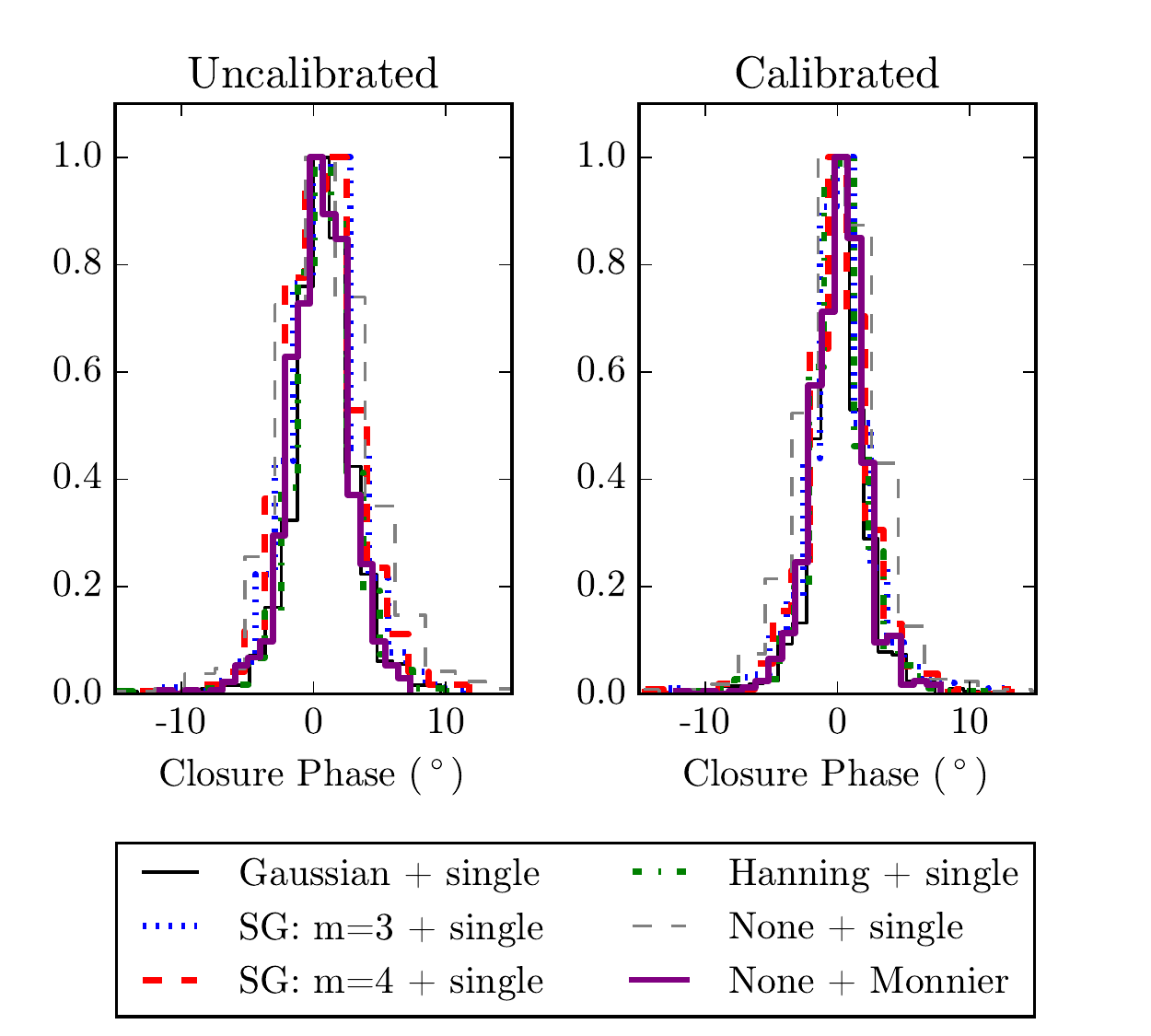}
\caption{Comparison of uncalibrated (left) and calibrated (right) closure phase scatters for different window functions and calculation methods.}
\label{fig-cpcomp}
\end{figure}

We use the scatter in the uncalibrated and calibrated data to compare the various closure phase calculation methods (see Figure \ref{fig-cpcomp}).
For both the calibrated and uncalibrated data, the single-pixel method without windowing results in the highest scatter ($\sigma_{uncal} = 3.8^\circ$, $\sigma_{cal} = 3.4^\circ$) and the ``Monnier" method without windowing results in the lowest scatter ($\sigma_{uncal} = 2.3^\circ$, $\sigma_{cal} = 2.1^\circ$). 
The single-pixel + window methods have a range of intermediate scatters ($\sigma_{uncal} = 2.4^\circ-3.0^\circ$, $\sigma_{cal} = 2.2^\circ-2.6^\circ$). 
Here, the narrower windows, which have wider Fourier transforms and thus farther-reaching inter-pixel correlations, lead to lower scatter in both the calibrated and uncalibrated data.

The ``Monnier" method is computationally more expensive than windowing, since hundreds or thousands of pixel triangles could exist for each closure phase.
However, since windowing blurs the Fourier transform, it could contaminate closure phases by incorporating phase information from spatial frequencies that do not form closing triangles.
For most datasets the ``Monnier" method also results in the lowest scatter.
For these reasons we use it over windowing for closure phase calculations.
The single-pixel + window method is only $\sim 0.1-0.2^\circ$ higher scatter and would be useful with limited computational resources or large datasets.

\section{Kernel Phase Projection}\label{sec-kps}

We project closure phases into linearly independent quantities called kernel phases \citep{2010ApJ...724..464M} so that we can fit uncorrelated observables.
To calculate kernel phases, we assume that our observed phase vector, $\mathbf{\Phi}$, can be written as a linear combination of instrumental phases, $\phi$, plus any intrinsic source phase $\mathbf{\Phi_0}$ (following the notation in \citet{2015ApJ...801...85S}):
\begin{equation}
\mathbf{\Phi} = \mathbf{A} \cdot \phi + \mathbf{\Phi_0}
\end{equation}
For a non-redundant mask, $\phi$ contains the instrumental phase measured on each of N sub-apertures. 
$\mathbf{A}$ is an M by N matrix that describes how those are combined to form the phase measured for each of $\mathrm{M} = {\mathrm{N} \choose \mathrm{2}}$ hole pairs.
We search for the kernel, or nullspace, of $\mathbf{A}$ so that the instrumental phase signal is eliminated, or
\begin{equation}
\mathbf{K} \cdot \mathbf{A} = \mathbf{0}.
\end{equation}

We find $\mathbf{K}$ using singular value decomposition of $\mathbf{A^T}$:
\begin{equation}
\mathbf{A^T} = \mathbf{U} \cdot \mathbf{W} \cdot \mathbf{V^T}.
\end{equation}
Here $\mathbf{U}$ is an $N\times N$ unitary matrix and $\mathbf{V}$ is an $M \times M$ unitary matrix.
$\mathbf{W}$ is an $N \times M$ diagonal matrix with zeros and positive values only.
The columns in $\mathbf{V}$ corresponding to the zero values in $\mathbf{W}$ form the null space of $\mathbf{A}$. 
These make up the rows of $\mathbf{K}$.

We already have a matrix $\mathbf{T}$, that is not linearly independent but that does eliminate instrumental phase signals.
$\mathbf{T}$ projects the M phases into ${\mathrm{N} \choose \mathrm{3}}$ closure phases:
\begin{equation}
\mathbf{\Phi_{CP}} = \mathbf{T}\cdot\mathbf{A}\cdot\phi + \mathbf{T}\cdot\mathbf{\Phi_0}.
\end{equation}
We can thus project the closure phases into kernel phases using the matrix $\mathbf{B}$ such that
\begin{equation}
\mathbf{B}\cdot\mathbf{T} = \mathbf{K}.
\end{equation}
Since $\mathbf{K}$ has full row rank, but not full column rank, it has a right inverse ($\mathbf{K^{-1}_R}$) only:
\begin{equation}
\mathbf{K}\cdot\mathbf{K^{-1}_{R}} = \mathbf{I} = \mathbf{B}\cdot\mathbf{T}\cdot\mathbf{K^{-1}_{R}}.
\end{equation}
$\mathbf{B}$ also only has a right inverse:
\begin{equation}
\mathbf{B^{-1}_{R}} = \mathbf{T}\cdot\mathbf{K^{-1}_{R}}.
\label{eq-binv}
\end{equation}
We then take the left inverse of Equation \ref{eq-binv} to calculate $\mathbf{B}$, and use it to project the closure phases into kernel phases.
We form the kernel phase variances by taking the diagonal values of the projected closure phase covariance matrix:
\begin{equation}
\mathbf{C_k} = \mathbf{B}\cdot\mathbf{C_{CP}}\cdot\mathbf{B^T}.
\end{equation}

We can form statistically independent kernel phases \citep{2013MNRAS.433.1718I} by applying another projection to diagonalize the kernel phase covariance matrix, $\mathbf{C_k}$. 
We diagonalize $\mathbf{C_k}$ by the spectral theorem:
\begin{equation}
\mathbf{C_k} = \mathbf{U}\cdot\mathbf{W}\cdot\mathbf{U^*},
\end{equation}
where $\mathbf{U}$ is a unitary matrix whose columns contain the eigenvectors of $\mathbf{C_k}$.
$\mathbf{K_s}$, a statistically independent kernel phase projection, is then calculated according to
\begin{equation}
\mathbf{K_s} = \mathbf{U}\cdot\mathbf{K}.
\end{equation}

\begin{figure}
\epsscale{1.0}
\plotone{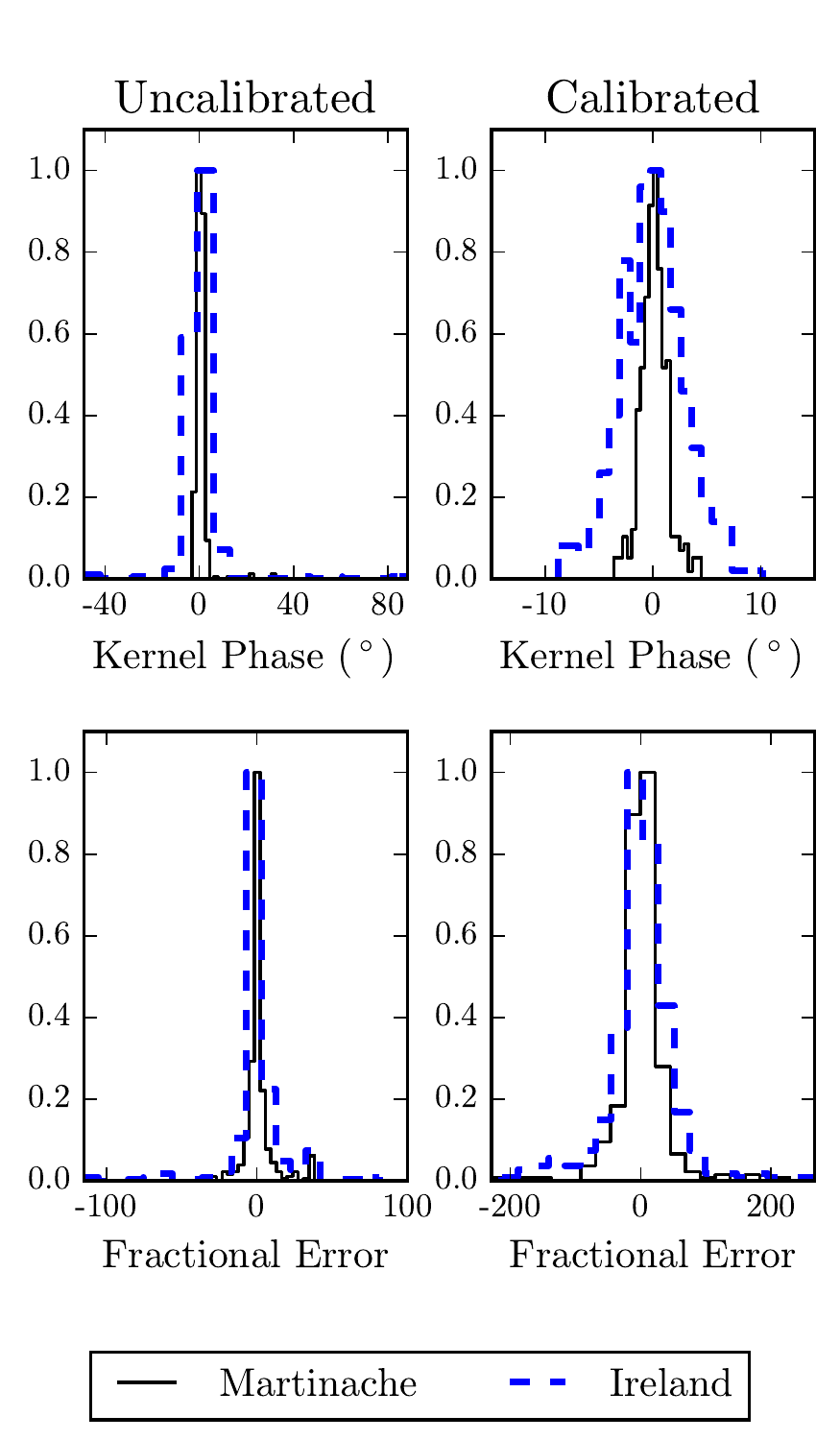}
\caption{Histograms of uncalibrated (left) and calibrated (right) kernel phases (top) and their fractional errors (bottom) for the ``Martinache"  (without $\mathbf{C_{CP}}$ diagonalization; solid black lines) and ``Ireland" (with $\mathbf{C_{CP}}$ diagonalization; dashed blue lines) projection methods.}
\label{fig-kpcomp}
\end{figure}

Figure \ref{fig-kpcomp} compares kernel phases calculated using $\mathbf{K}$ (the ``Martinache" projection) and $\mathbf{K_s}$ (the ``Ireland" projection).
The top two panels show uncalibrated and calibrated observations of two unresolved stars from December 2014. 
The Martinache kernel phases have lower calibrated and uncalibrated scatters than the Ireland kernel phases.
The two projections may have relative scalings, which would result in kernel phase signals of different magnitudes for a given source morphology.
To ensure a fair comparison we also show their fractional errors (see Figure \ref{fig-kpcomp}, bottom panels). 
While the Ireland projection has only slightly higher fractional error for the majority of the kernel phases, it has many more outliers than the Martinache projection.
We thus use the Martinache projection, which incorporates information only about the mask and not about the observations, for model fits to kernel phases.

\section{Squared Visibility Generation}

We calculate squared visibilities by summing the power ($\left| A e^{i \phi}\right|^2$) in all pixels corresponding to each baseline in the complex visibilities.
We measure any bias by taking the average power for regions in Fourier space without signal.
We subtract this bias and then normalize by the power at zero baseline (equivalent to normalizing by total power in the interferogram).

\begin{figure}
\epsscale{1.0}
\plotone{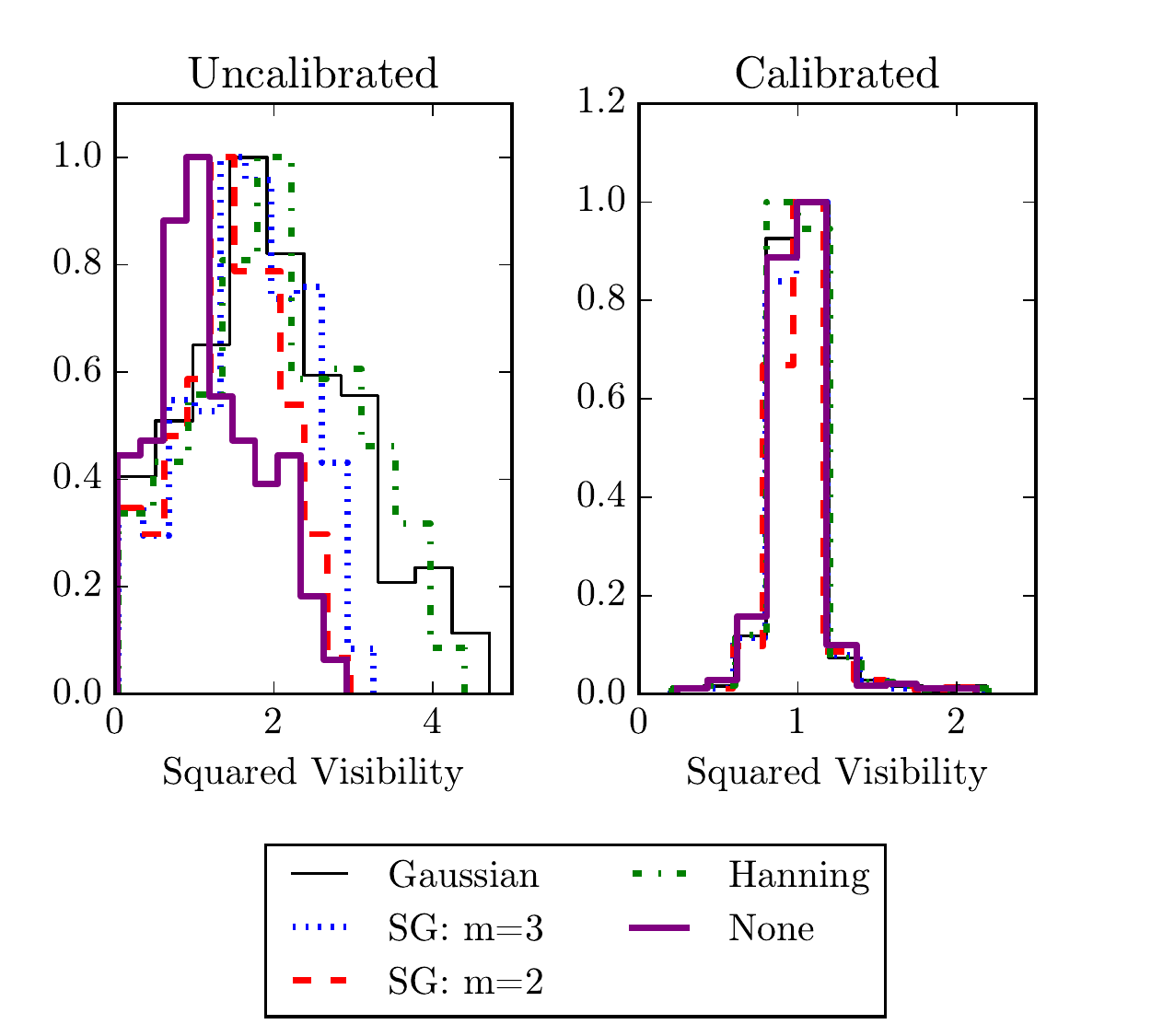}
\caption{Histograms of uncalibrated (left) and calibrated (right) squared visibilities for different window functions.}
\label{fig-v2comp}
\end{figure}

Figure \ref{fig-v2comp} compares squared visibilities generated after applying the different window functions. 
The choice of window function has a more dramatic effect on the uncalibrated visibilities than on the calibrated visibilities. 
Windowed squared visibilities have much higher values before calibration.
This is because convolving in Fourier space decreases the power at a spatial frequency of zero relative to the mask spatial frequencies.
However this effect is uniform between the target and calibrator observations and thus calibrates out.

We note that this squared visibility comparison is for a dataset with very good sky subtraction; the edges of the subframed images have an average pixel value very close to zero.
However, in conditions where sky subtraction is difficult, average pixel values may deviate from zero at the edges of the subframed images.
This can cause ringing in the Fourier plane that would contaminate the squared visibilities.
Here, windowing helps to remove this noise, which may not be consistent between target and calibrator observations in variable conditions.\\\\

\section{Calibration}

\subsection{Instrumental Signal Fitting}

\begin{figure}
\epsscale{1.0}
\plotone{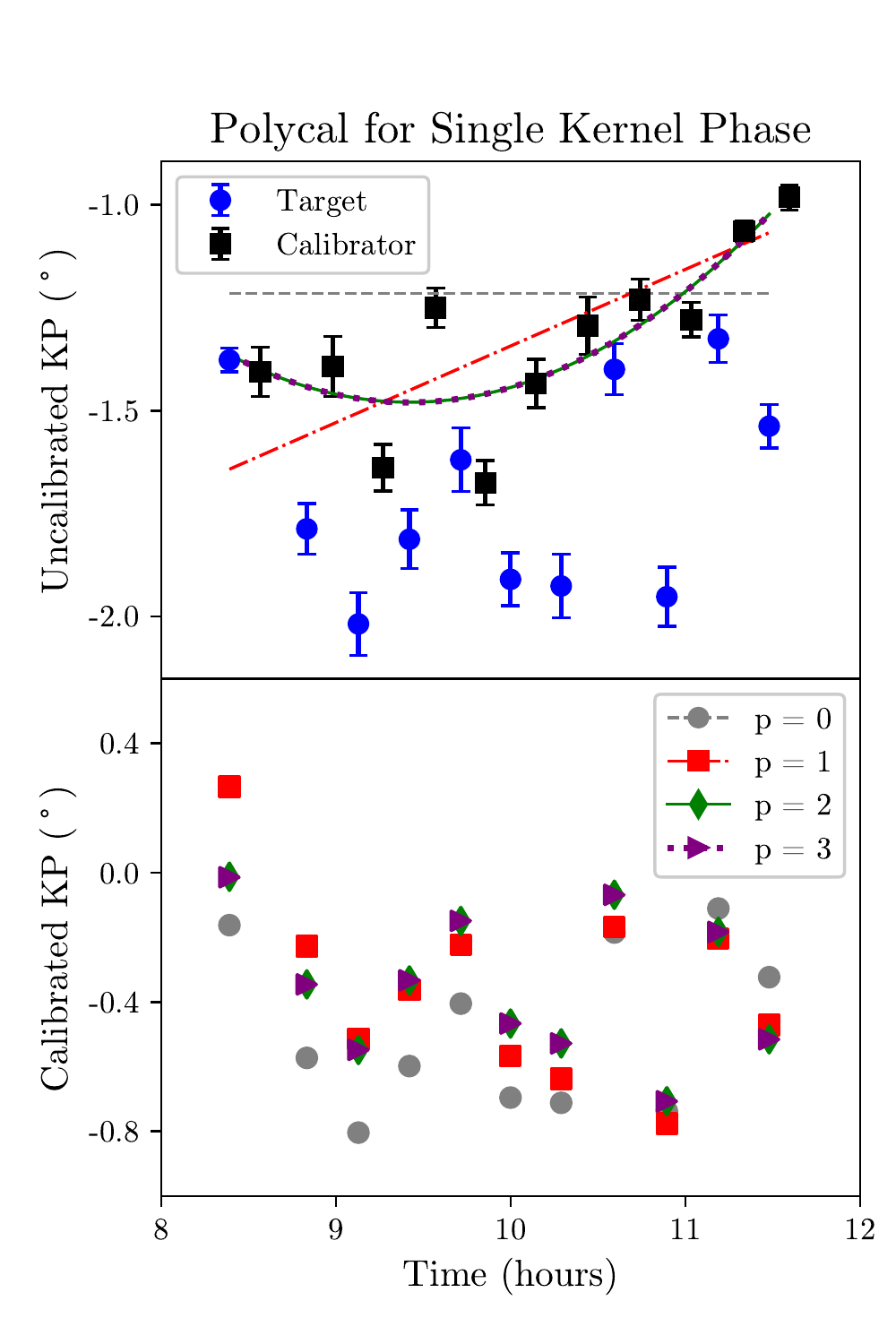}
\caption{Polycal example for a single kernel phase. The top panel shows the calibrator (black points) and target (blue points) observations for a single kernel phase as a function of time. The plotted lines show polynomial fits as a function of time with different maximum orders. The bottom panel shows the calibrated kernel phases over time, with different colors indicating fits up to different orders in time. Since the best fit 3rd order polynomial was nearly identical to the best fit 2nd order polynomial, the purple triangles and purple dotted line are plotted over the green diamonds and green solid line. Here a 2nd order polynomial was eventually chosen, since it led to the lowest scatter in the calibrated observables.}
\label{fig-polyex}
\end{figure}

\begin{figure}
\epsscale{1.0}
\plotone{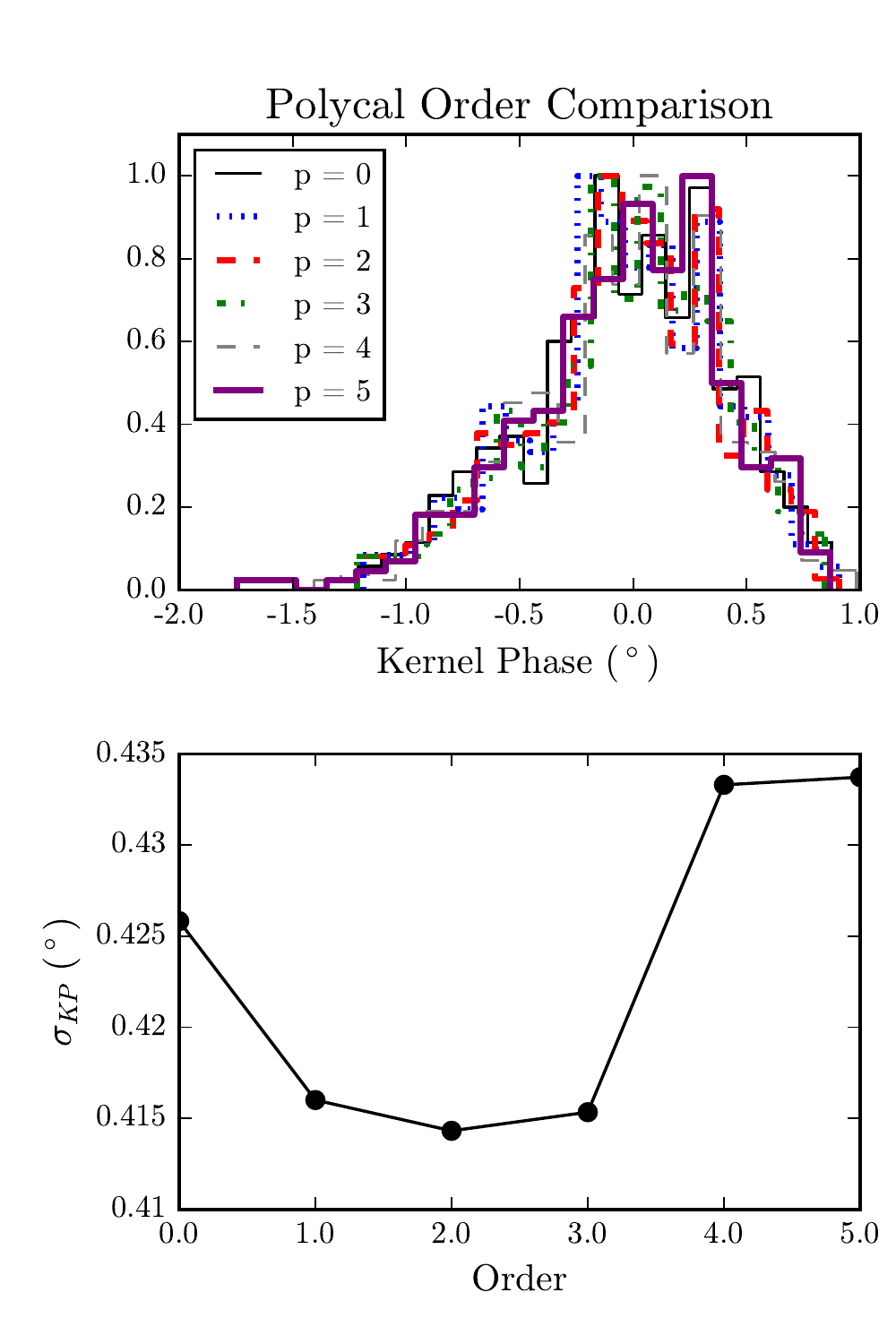}
\caption{Comparison of calibrated kernel phases for Polycal. The top panel shows histograms for an entire night of calibrated kernel phases for different maximum order polynomials. The bottom panel shows the standard deviation of a night of calibrated kernel phases as a function of maximum order.}
\label{fig-polyselect}
\end{figure}

We calibrate the phases and squared visibilities in two different ways.
In the first, which we refer to as Polycal, we fit polynomial functions in time to the calibrator observations. 
A zeroth order function in time would represent a constant instrumental signal, while a very high order polynomial would indicate a highly variable instrumental signal.
We calculate a different set of coefficients for each kernel phase as a function of time.
We set the maximum allowed order to be the same for all kernel phases, choosing the one that minimizes the scatter in all of the calibrated observables without over-fitting the calibrator measurements.
We sample this polynomial function at the time of the target observations to find the instrumental signal.
We subtract the instrumental signal from the raw kernel and closure phases and divide it into the raw squared visibilities.
Figure \ref{fig-polyex} shows an example calibration for a single, linearly independent kernel phase (see Section \ref{sec-kps}).
Figure \ref{fig-polyselect} compares the calibrated kernel phases for polynomials with terms up to different orders in time, and shows how the observed scatter changes with maximum order.

\begin{figure}
\epsscale{1.0}
\plotone{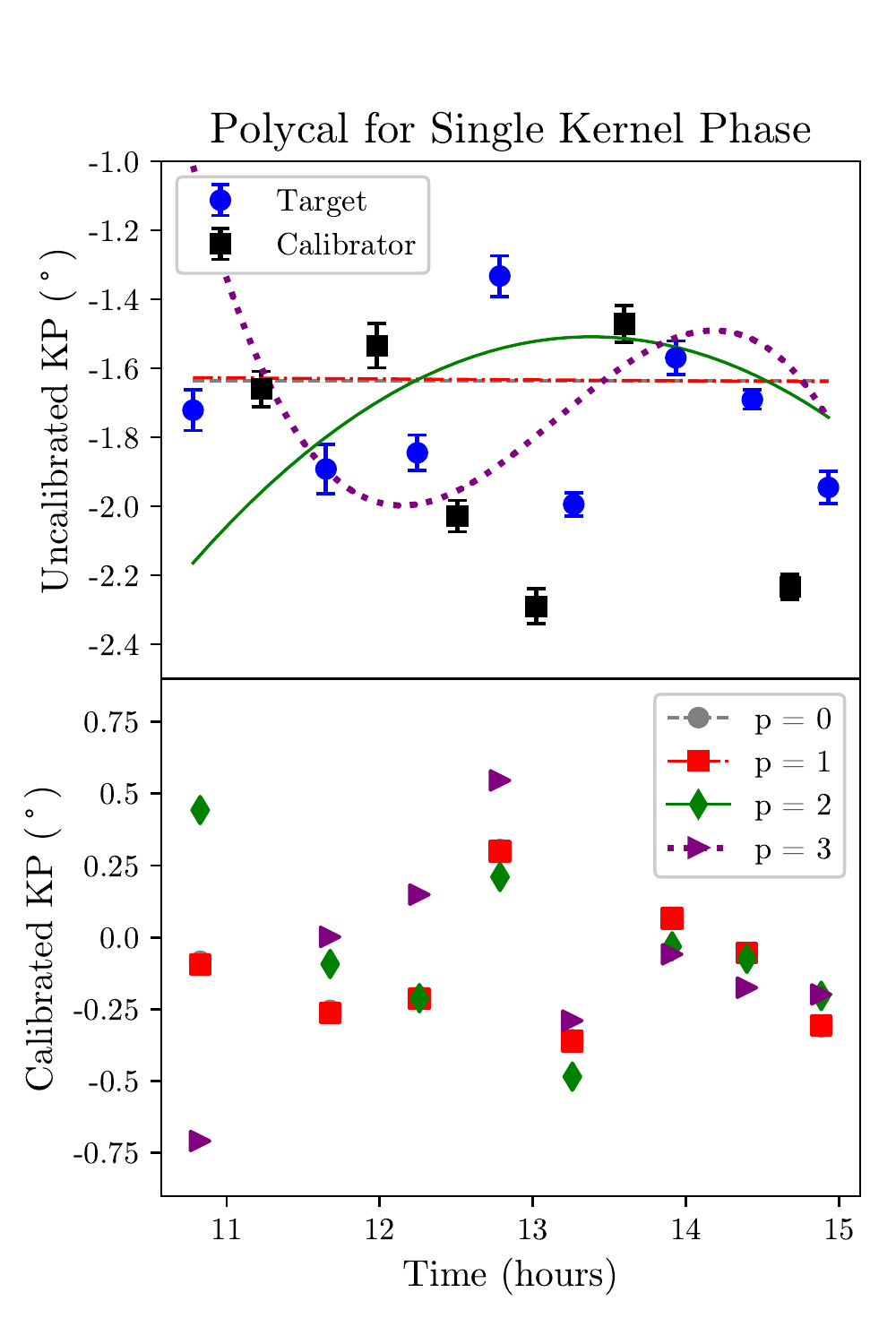}
\caption{Polycal example for a single kernel phase. The top panel shows the calibrator (black points) and target (blue points) observations for a single kernel phase as a function of time. The plotted lines show polynomial fits as a function of time with different maximum orders. The bottom panel shows the calibrated kernel phases over time, with different colors indicating fits up to different orders in time. Since the best fit 1st order polynomial was nearly identical to the best fit 0th order polynomial, the red squares and red dot-dashed line are plotted over the grey circles and grey dashed line. Here a 2nd order polynomial was eventually chosen, since it led to the lowest scatter in the calibrated observables.}
\label{fig-polyexb}
\end{figure}

Polycal is best performed when calibrator observations can be taken before the first and after the last target observation.
Figure \ref{fig-polyexb} illustrates the potential issues that can occur when calibrator observations do not bookend the target observations.
For this dataset, the large scatter in the calibrator observations leads to best-fit polynomials that prefer large coefficients for higher order terms. 
These high order polynomials match the first calibrator observation well, but have unrealistic values at the earlier time of the first target observation (see purple dotted line and solid green line in Figure \ref{fig-polyexb}).
Second order polynomials often provide calibration that is comparable to or better than higher orders.
They also tend to avoid these issues when target observations are not bookended by calibrator measurements.

\subsection{Optimized Calibrator Weighting}

The second calibration method is an optimized calibrator weighting detailed in \citet{2012ApJ...745....5K} and \citet{2013MNRAS.433.1718I}, similar to the ``locally optimized combination of images" \citep[LOCI; ][]{2007ApJ...660..770L} algorithm applied in filled-aperture direct imaging.
Here, calibrator weights are first chosen to minimize the $\chi^2$ of the null model (zero phase signal).
They are then chosen iteratively to minimize the $\chi^2$ of the global best fit model until the best fit converges. 
Following the notation in \citet{2013MNRAS.433.1718I}, for each uncalibrated set of kernel phases $\mathbf{x_t}$, we search for a set of N weights ($a_k$, where $k$ goes from 1 to N), to average the N calibrator measurements.
The calibrated kernel phases for one pointing ($\mathbf{x_c}$) are:
\begin{equation}
\mathbf{x_c} = \mathbf{x_t} - \sum_{k=1}^{N} a_k \mathbf{x_k}.
\end{equation}
The $\chi^2$ of the null model is then
\begin{equation}
\chi^2 = \sum \frac{\mathbf{x_c}}{\mathbf{\sigma_c}^2},
\end{equation}
where $\sigma_c$ are the calibrated kernel phase errors, given by
\begin{equation}
\mathbf{\sigma_c}^2 = \mathbf{\sigma_t}^2 + \sum_{k=1}^{N} a_k  \mathbf{\sigma_k}^2.
\end{equation}

Several different likelihood functions can be maximized to find the optimal calibrator weights.
The simplest likelihood function would be
\begin{equation}
L = \exp\left(-0.5~\chi^2\right).
\end{equation}
However, maximizing this likelihood function could wash out true signals.
To try to prevent this, \citet{2012ApJ...745....5K} add a regularizer, $\pi$, to the likelihood function.
Arbitrary functions can be used for $\pi$; the calibration used in \citet{2012ApJ...745....5K} applies the following regularization:
\begin{equation}
\pi = \exp\left(-0.5~a_k^2 \sum \frac{\mathbf{\sigma_k}^2}{\mathbf{\sigma_t}^2}\right),
\end{equation}
which punishes large weights and calibrator measurements with large errors.
The likelihood function is then:
\begin{equation}
L = \exp\left(-0.5~\chi^2\right) \times \pi
\label{eq-loci}
\end{equation}
This calibration is performed iteratively to constrain an additional ``calibration error" term, $\Delta$, intended to account for errors beyond the random component found in $\sigma_t$ and $\sigma_k$.
The $\Delta$ term is a constant added to all errors (calibrators and target), designed so that the reduced $\chi^2$ of the calibrated kernel phases is equal to 1:
\begin{equation}
\chi^2_r = \frac{1}{N} \sum \frac{\mathbf{x_c}}{\mathbf{\sigma_c}^2} = 1
\end{equation}

\begin{figure}
\epsscale{1.2}
\plotone{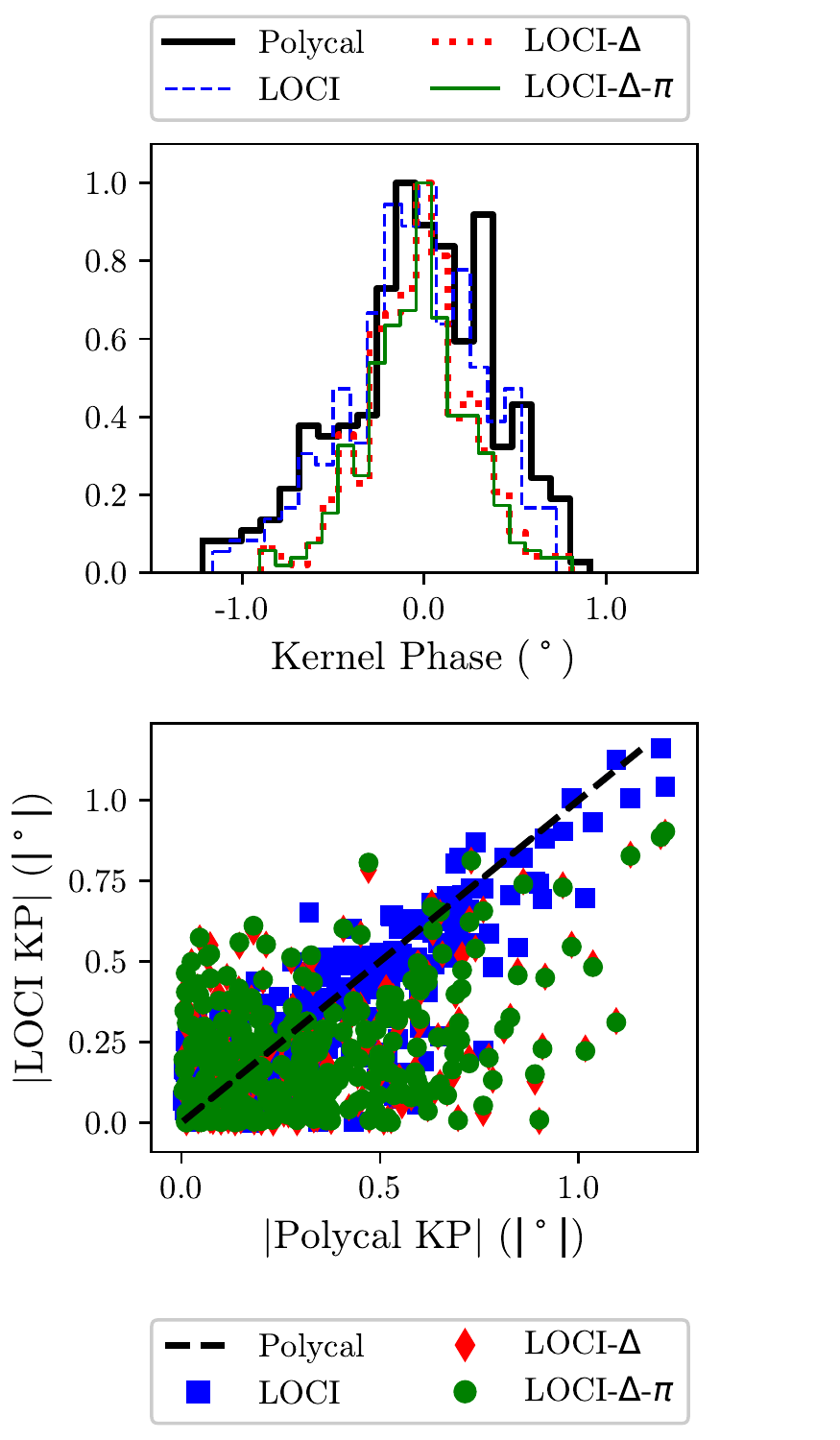}
\caption{LOCI regularization comparison. In the top panel, the black line shows the Polycal kernel phase distribution for Dataset 2 from Figure \ref{fig-loci}. The dashed blue line shows the LOCI calibration from Figure \ref{fig-loci}, the red dotted line a LOCI calibration with no ``calibration error" term, and the green solid line a LOCI calibration with no regularization or  ``calibration error" term. The bottom panel shows the three LOCI calibrations' absolute value kernel phases plotted against those for the Polycal calibration.}
\label{fig-locireg}
\end{figure}

Figure \ref{fig-locireg} demonstrates the ability of the regularizer, $\pi$ and the calibration error term $\Delta$ to bias the calibration.
The blue line / points show the LOCI calibrated kernel phases for Dataset 1 in Figure \ref{fig-loci}.
The red line / points show the LOCI calibration without using the calibration error term, and the green line / points show the LOCI calibration without using $\Delta$ or $\pi$.
The discrepancy in the overall distribution of kernel phases (top panel), as well as the in the values for the individual kernel phases (bottom panel), show that the calibration can change significantly depending on the choice of error scaling and regularizer.
With no error scaling or regularizer, the LOCI-like calibration could eliminate all astrophysical signal as the number of calibrator measurements becomes large.
Thus care must be taken in choosing an appropriate regularizer.\\\\

\subsection{Calibration Comparison}

\begin{figure*}
\epsscale{0.75}
\plotone{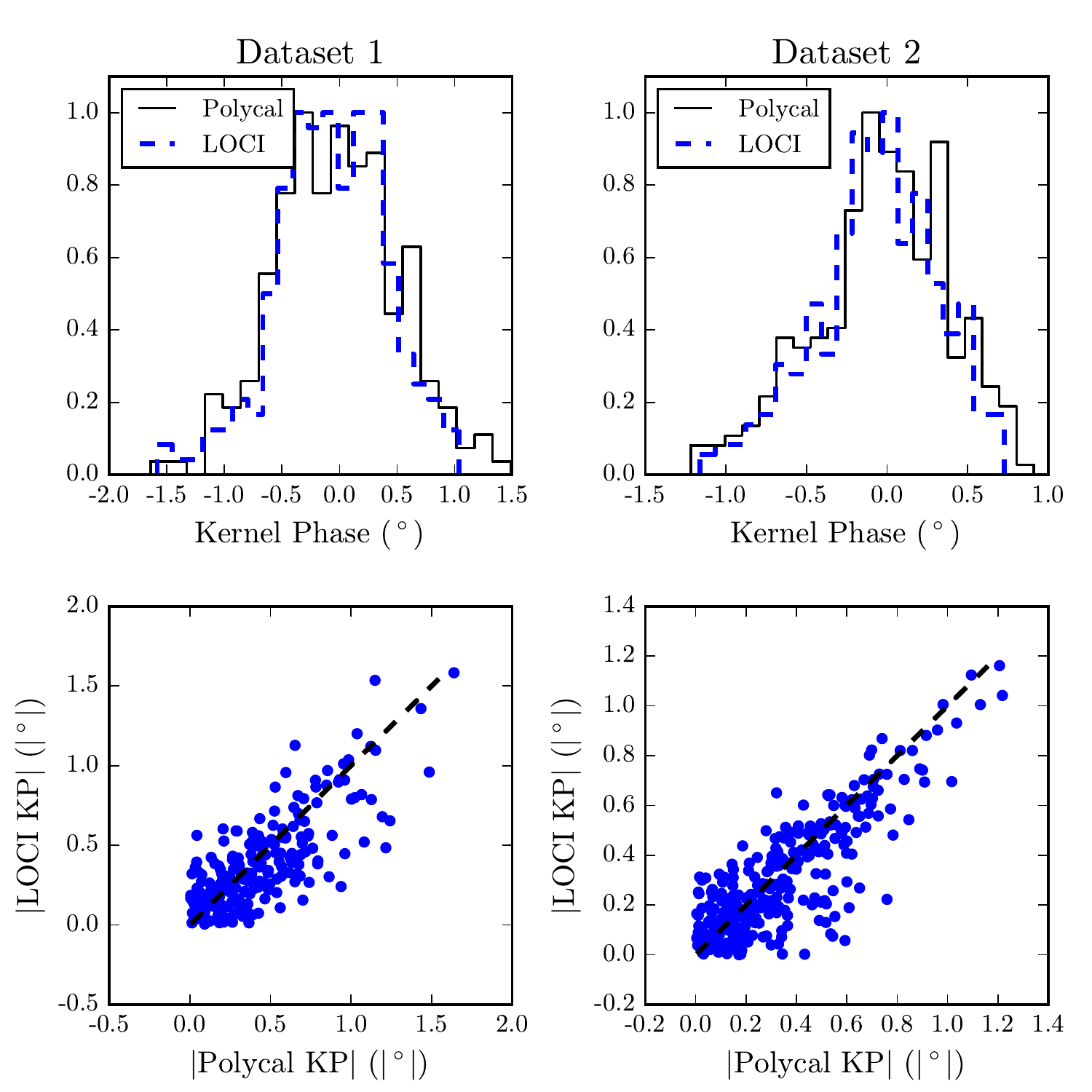}
\caption{Comparison of Polycal and LOCI-like calibrated kernel phases for two nights of data. The top two panels show the histograms of calibrated kernel phases for Polycal in black and LOCI in blue. The bottom two panels show the absolute value of the LOCI kernel phases plotted against those for the Polycal calibration.  The dashed line shows a 1:1 correlation.}
\label{fig-loci}
\end{figure*}

Figure \ref{fig-loci} compares the Polycal kernel phases to the LOCI-like kernel phases using the likelihood function defined by Equation \ref{eq-loci} for two different datasets.
The histograms in the top two panels show that the LOCI-like calibration does not decrease the scatter much more than the Polycal calibration.
The bottom two panels show the absolute value of the LOCI kernel phases plotted against the absolute value of the Polycal kernel phases. 
Deviations from the dashed line in these panels (indicating a 1:1 relationship) show differences between the two calibration methods for each kernel phase.
While LOCI does not change the distribution of all kernel phases significantly, it does change the individual kernel phase values.

To compare the LOCI and Polycal calibrations under different noise conditions, we injected binary signals into L band observations of two unresolved calibrator stars in two datasets.
The mean uncalibrated kernel phase scatter for the two datasets was 1.1$^\circ$ and 1.5$^\circ$, leading to calibrated scatters of 0.3$^\circ$ and 0.6$^\circ$, respectively, using Polycal.
We first checked for both sets of observations that no significant companion signals existed, by calibrating each star using the other and fitting companion models to the calibrated data.
We then chose one star as a mock target in which to inject companion signals, and used the other to calibrate it using both methods. 
We fit the Polycal kernel phases a single time using the open source Markov-Chain Monte Carlo package \texttt{emcee} \citep{2013PASP..125..306F}.
For LOCI we calibrated the kernel phases iteratively, calibrating toward the global best fit model until the fit converged.

We injected fake signals at a separation of 80 mas with contrasts ranging from $\Delta_L = 1$ to $\Delta_L = 7$.
Polycal could recover slightly higher contrast companions than LOCI, since the optimized calibration washed out faint input signals.
LOCI could not recover companions with contrasts fainter than $\Delta_L \sim 5.5$ mag and $\Delta_L \sim 7$ mag for uncalibrated kernel phase scatters of 1.5$^\circ$ and $1.1^\circ$, respectively.
Here, the initial LOCI calibration left behind a few noise spikes with likelihoods comparable to the input companion model's.
In this scenario it would be possible to calibrate toward the wrong companion signal by selecting a noise spike rather than the true signal.
This highlights the importance of quantifying Type 1 and Type 2 errors in NRM companion modeling \citep[e.g.][]{2015ApJ...801...85S}.
At these contrasts, Polycal recovers the input signal in both datasets, although the parameter uncertainties are large.
For brighter sources, both calibrations recover the true signal, resulting in comparable fit parameter errors for the $1.5^\circ$ scatter dataset.
With lower scatter ($1.1^\circ$), the parameter uncertainties from the LOCI calibration were $\sim 2$ times larger.

The injected companion detections and kernel phase scatter comparison show that Polycal can provide comparable calibration to LOCI without regularization and error scaling. 
For at least the case of a simple binary, Polycal results in similar or lower parameter uncertainties and recovers high contrast companion signals more reliably.
Furthermore, in this implementation the optimized calibration takes the science observations into account; it thus cannot be a completely unbiased measurement of the instrumental signal.
This could be avoided by observing many calibrators and, for each target observation, computing an optimal weighting that minimizes the calibrator signal closest in time to the target observation.
However this is nearly identical to Polycal.
Since Polycal depends only on the calibrator observations, requires no regularization or error scaling terms, and performs comparably to LOCI, for inexperienced users we recommend using simple calibrations like this over the optimized calibrator weighting.

We note that using negative calibrator weights is possible only with the LOCI calibration.
This is important if the target and calibrator spectra differ, since dispersion can cause the target instrumental kernel phases to be best represented by the difference in calibrator kernel phases \citep[e.g.][]{2013MNRAS.433.1718I}.
When uncorrected, dispersion can mimic a close-in companion \citep[e.g.][]{2012A&A...541A..89L}.
We avoid this by choosing calibrators whose fluxes are close to the target at both the wavefront sensing and science wavelengths.
In situations where this is not possible, the optimized calibrator weighting may yield better results.

\section{Image Reconstruction}
Uneven and incomplete Fourier coverage, analogous to imaging in radio interferometry \citep[e.g.][]{1974A&AS...15..417H}, makes image reconstruction an ill-posed problem.
With perfect knowledge of the complex visibilities, synthesizing an image would only require an inverse Fourier transform.
However, the number of pixels in a reconstructed image is much larger than the number of Fourier phases and amplitudes that we can constrain.
Furthermore, since the Earth's atmosphere corrupts the individual phases, we measure combinations of phases that are robust to atmospheric errors.
As a result, we cannot independently measure all of the phases for the array and have less phase information than in the radio case. 
This lack of Fourier information means that an infinite number of model images can provide comparable fits to the observations.
To reconstruct images from NRM observations, we thus use algorithms that maximize the likelihood of the data while also satisfying a regularization constraint \citep[e.g][]{2010ISPM...27...97T}.
Regularizers, which were first used to compensate for incomplete Fourier coverage in the radio \citep[e.g.][]{1974A&AS...15..417H,1985A&A...144..381T,1974A&AS...15..383A}, are chosen by hand to impose prior knowledge such as positivity, smoothness, sharp edges, or sparsity \citep[e.g.][]{2011A&A...533A..64R}. \\

\subsection{Optimization Engines}
While many different optimization engines exist for finding the best image, they can be split into two general categories.
Deterministic algorithms take steps proportional to the gradient of the likelihood function at the current image state. 
These include the steepest descent method in the Building Block Method \citep[e.g.][]{1993SPIE.1983..203H}, the constrained semi-Newton method \citep[used in the algorithm MiRA; e.g.][]{2008SPIE.7013E..1IT}, and the trust region method \citep[used in the algorithm BSMEM; e.g.][]{1994IAUS..158...91B,2008SPIE.7013E..3XB}.
Stochastic algorithms, on the other hand, involve moving flux elements randomly in the image plane during iterations in Monte Carlo Markov Chains (MCMC).
One stochastic algorithm is MACIM \citep{2006SPIE.6268E..1TI}, a simulated annealing method that accepts or rejects images based on a temperature parameter.
A newer stochastic algorithm is SQUEEZE \citep{2010SPIE.7734E..2IB}, which can perform parallel simulated annealing, where the results of several simulated annealing chains are averaged.
SQUEEZE can also perform parallel tempering, where several MCMC chains at different temperatures can exchange information as they run.

\subsection{Regularizers}
Many regularizers can be used to constrain the image reconstruction in different ways.
Some regularizers punish excursions from a prior image, or a default image in the absence of any prior information. 
An example is the Maximum Entropy, or MEM, regularization, which favors the least-informative reconstruction by maximizing an entropy term in the likelihood function \citep[e.g.][]{1987Natur.328..694H,1994IAUS..158...91B}.
The results of regularizers such as MEM depend heavily on the choice of prior image \citep[e.g.][]{2016ASSL..439...75B}.

Compressed sensing regularizations decompose the image into a linear combination of basis functions and then apply constraints.
A simple example would be sparsity in the pixel basis.
This was first enforced by the Building Block Method \citep{1993SPIE.1983..203H} and can result in large areas with low flux.
Sparsity can also be imposed in the gradient of the image, which would preserve uniform flux and edges \citep[e.g.][]{2011A&A...533A..64R}.
A more recent regularization that can be applied in SQUEEZE is one that imposes sparsity after decomposing the image into a wavelet basis \citep[e.g.][]{2010SPIE.7734E..2IB}.
The wide variety of available regularizers, of which these are just a few examples, shows that care must be taken in choosing one and in understanding its effect on the final reconstructed image.

\subsection{Degeneracies}\label{sec:degen}

Comparing the reconstructed images from LBT NRM observations of MWC 349A \citep{2017ApJ...844...22S} and LkCa 15 \citep{2015Natur.527..342S} illustrates the effects of prior choice and array configuration on the final reconstructed images.
The MWC 349A dataset has two pointings with $\sim13^\circ$ change in parallactic angle, while the LkCa 15 dataset has 15 pointings with parallactic angles between $-65^\circ$ and $65^\circ$.
The LkCa 15 observations were taken with the LBT in single-aperture mode, using only baselines within each of the two primary mirrors. 
The MWC 349A data were taken in dual-aperture mode, but due to the small amount of sky rotation these data only have triple the single aperture resolution for a small range of position angles.

\begin{figure}
\epsscale{0.7}
\plotone{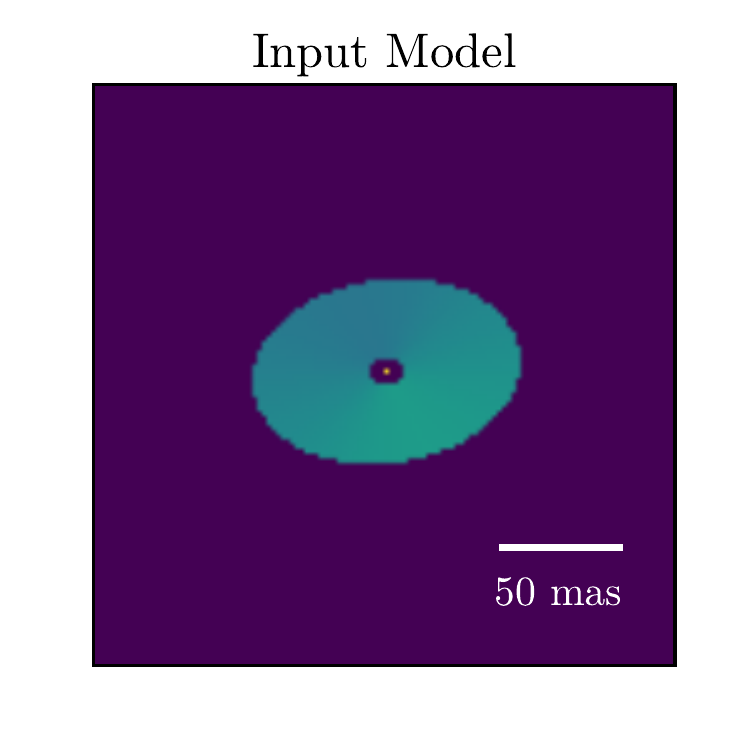}
\caption{Best fit skewed ring plus delta function model for observations of MWC 349A published in \citet{2017ApJ...844...22S}. This image was used as the input for the simulations shown in the bottom row of Figure \ref{fig-mwc349_prior} and in Figure \ref{fig-mwc349_sims}.}
\label{fig-mwc349_input}
\end{figure}

Figure \ref{fig-mwc349_input} shows the best fit skewed ring + delta function model for the MWC 349A observations. 
This model was determined by fitting the complex visibility data directly, and this does not depend on image reconstruction.

\begin{figure}
\epsscale{1.0}
\plotone{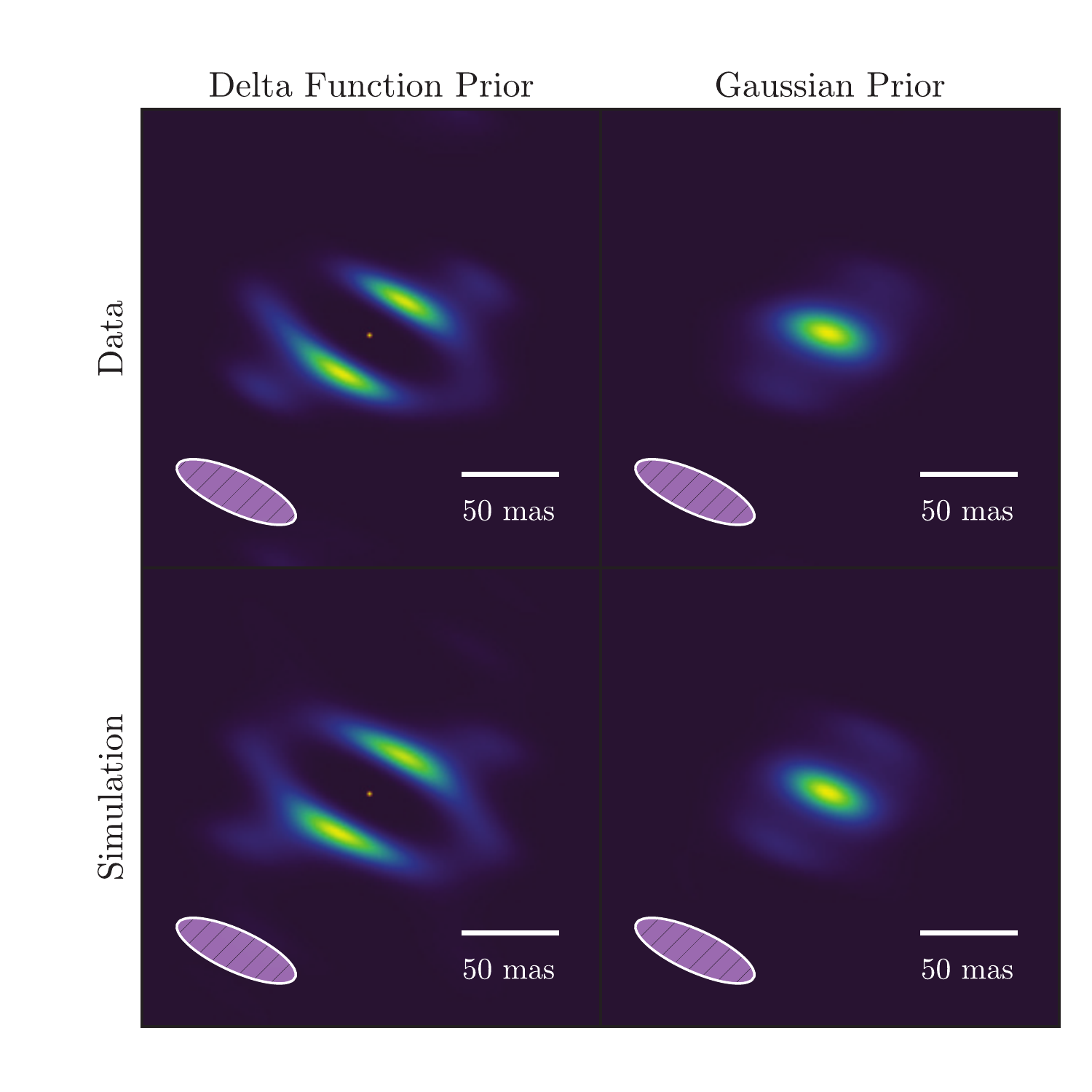}
\caption{From \citet{2017ApJ...844...22S}, reconstructed images from observations of MWC 349A (top row) and simulated observations of the best-fit skewed disk + delta function model (bottom row; see Figure \ref{fig-mwc349_input}). The left and right columns show reconstructions for delta function and Gaussian priors, respectively}
\label{fig-mwc349_prior}
\end{figure}

The top row of Figure \ref{fig-mwc349_prior} shows images reconstructed using BSMEM for the data, and the bottom row for simulated observations of the model image in Figure \ref{fig-mwc349_input}. 
The left and right columns show images reconstructed using delta function and Gaussian priors.
The simulations show that observations of the source shown in Figure \ref{fig-mwc349_input} lead to different reconstructed images depending on the choice of prior.
The delta function prior results in a single, bright pixel, surrounded by a dark hole roughly the same size as the central part of the synthesized beam (contours in Figure \ref{fig-mwc349_prior}). 
The Gaussian prior puts the same fractional flux into the central part of the beam as there is in the single pixel at the center of the delta prior reconstruction. 

\begin{figure*}
\epsscale{1.0}
\plotone{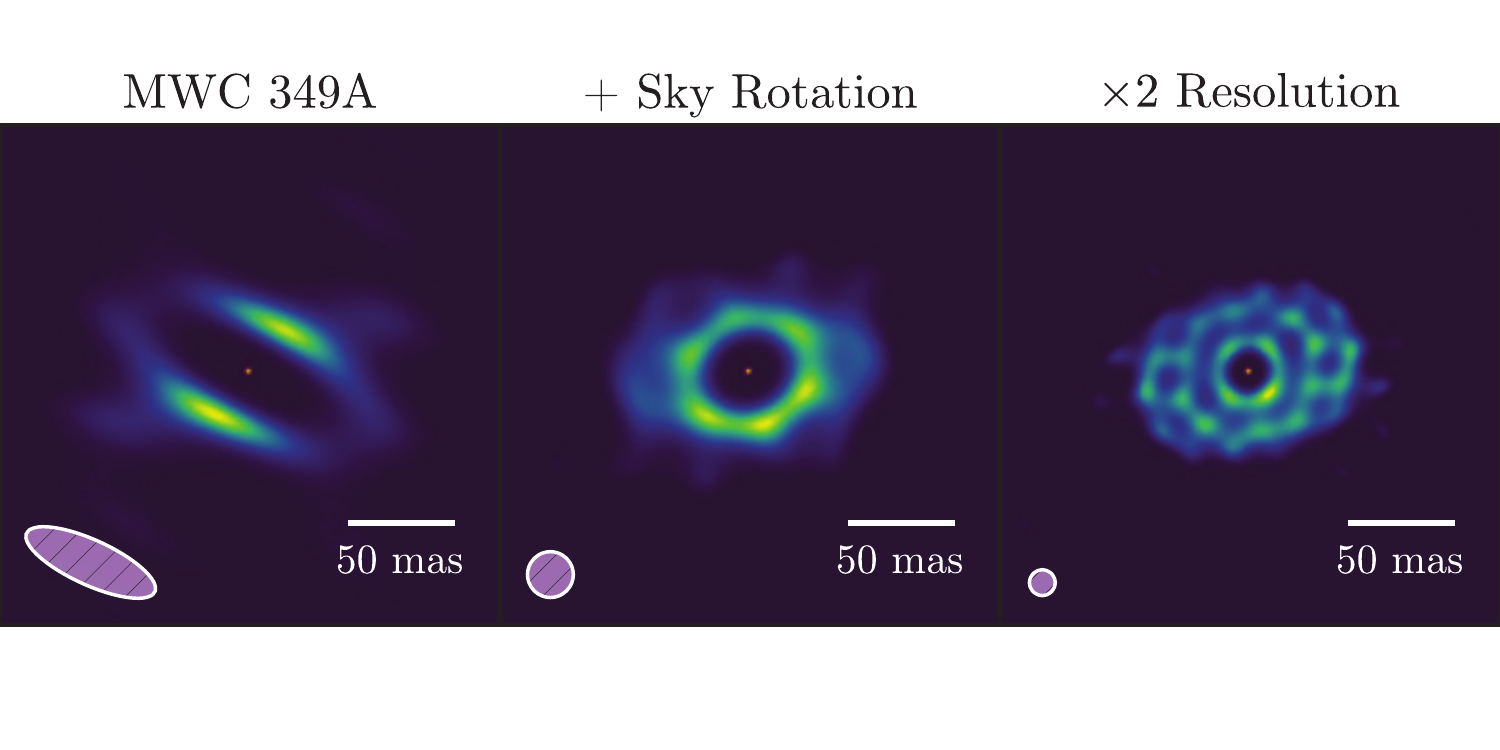}
\caption{Simulated observations of the model shown in Figure \ref{fig-mwc349_input}. The leftmost panel shows the MWC 349A ($u,v$) coverage and sky rotation. The center panel shows the same mask configuration but with very dense sky rotation, and the right panel shows the dense sky rotation case with a mask having twice the resolution.}
\label{fig-mwc349_sims}
\end{figure*}

\begin{figure}
\epsscale{1.0}
\plotone{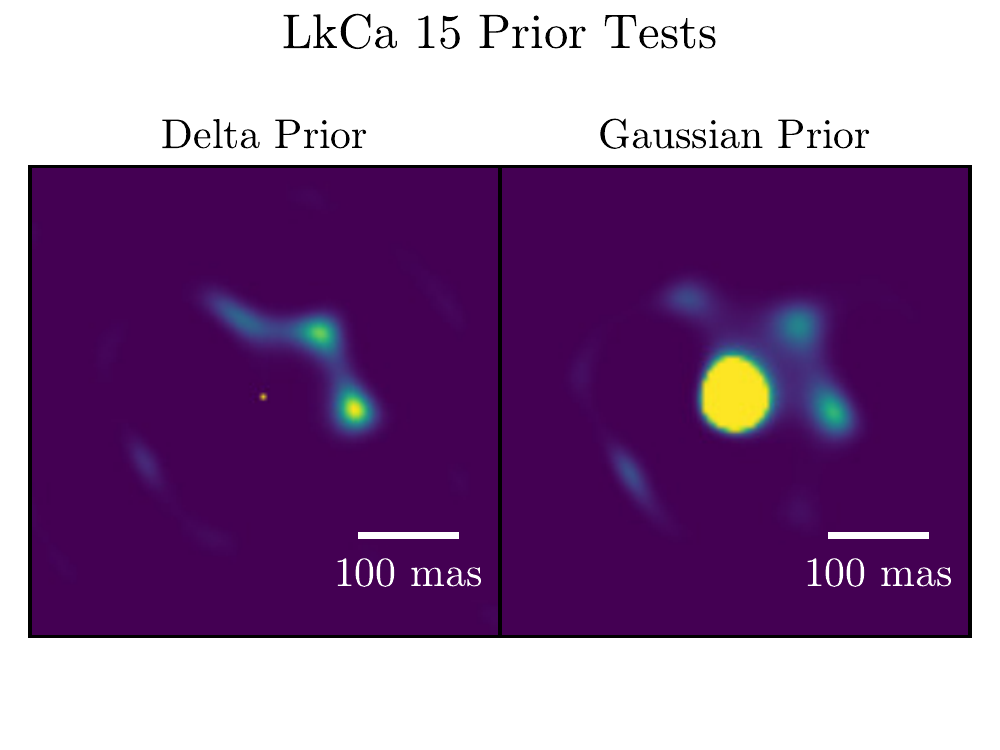}
\caption{Reconstructed images from the L$'$ LBT NRM dataset published in \citet{2015Natur.527..342S}. The left panel shows the results of using a delta function prior, and the right panel a Gaussian prior. The two images are shown on the same color scale.}
\label{fig-lkca15}
\end{figure}

Figure \ref{fig-mwc349_sims} shows the effects of improving the sky rotation (center panel), and resolution (right panel) of the observations.
Reconstructions improve with more complete Fourier coverage and longer baselines, showing the true extent and position angle of the input model despite an inadequate prior.
The LkCa 15 reconstructions (see Figure \ref{fig-lkca15}), illustrate this as well.
Again, the Gaussian prior results in a region of emission roughly the size of the beam that has the same fractional flux as the central pixel in the delta function reconstruction.
However here the sources are at large enough separation (compared to the array resolution) that their portion of the image is less affected by the choice of prior.
These degeneracies and their different effects on different datasets highlight the need for modeling in interpreting reconstructed images.

\subsection{Simulated Data Reconstructions: BSMEM versus SQUEEZE}\label{sec:sims}
Since choices made during the image reconstruction process can change the final image, we use simulated data, rather than observations, to compare the reconstructed images to eight different source brightness distributions.
The point of these reconstruction tests is to evaluate each algorithm's performance with no prior knowledge of the source morphology.
Thus we run each algorithm in identical configurations regardless of the input model. 
The detailed implementation and results of these simulations can be found in the Appendix.
We compare two image reconstruction algorithms, BSMEM \citep{1994IAUS..158...91B,2008SPIE.7013E..3XB}, and SQUEEZE \citep{2010SPIE.7734E..2IB}, that are often used on NRM datasets.
BSMEM is a deterministic algorithm with a MEM regularizer, while SQUEEZE is a stochastic algorithm with a variety of available regularizers.

With no knowledge of the source morphology, and with the potential for poor error bar estimation, BSMEM performs better than SQUEEZE.
The BSMEM reconstructions vary less when the data error bars are over- or under-estimated, and also when the relative closure-phase and squared visibility errors vary. 
Weighting the data by baseline length changes the results of both algorithms, with SQUEEZE providing better reconstructions of extended sources than BSMEM when long baselines were downweighted. 
Both BSMEM and SQUEEZE fail to converge when certain initial images are used for some input models (delta function initial images for BSMEM, and Gaussian initial images for SQUEEZE).
Neither algorithm behaves so poorly that the source morphology is completely unrecognizable, but recovering features like close-separation point sources is more difficult with SQUEEZE under certain conditions.

We note that, with prior knowledge of the source morphology, the relative performances of BSMEM and SQUEEZE may change as more optimal regularizations can be applied.
For example, using BSMEM with a delta function prior may be optimal when the target is expected to be a collection of point sources.
When large areas of extended emission are expected, using SQUEEZE with a weighting scheme that upweights short baselines may yield the best results.

\subsection{NRM on GMT: Comparison with LBTI}
GMT's expected performance makes non-redundant masking appealing for imaging at small angular separations.
The expected wavefront error due to imperfect segment phasing is $\sim50$ nm rms for bright guide stars \citep[e.g.][]{2016SPIE.9906E..6DQ}. 
Furthermore, while next generation adaptive optics systems are designed to reach rms wavefront errors less than $\sim100$ nm, their measured wavefront errors are $\sim100-150$ nm \citep[e.g.][]{2014PNAS..11112661M}. 
These wavefront errors, which would make traditional high resolution imaging difficult, can be characterized and calibrated using a non-redundant mask. 

\begin{figure}
\epsscale{1.0}
\plotone{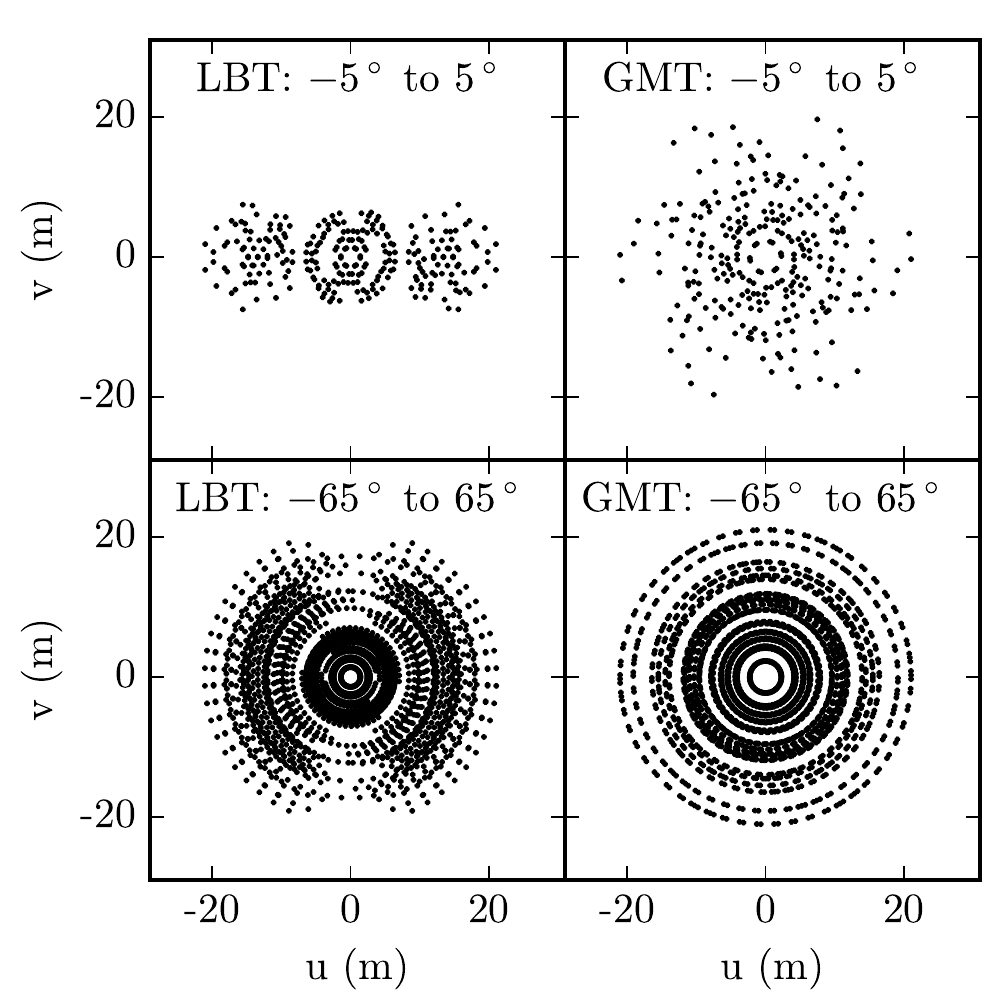}
\caption{Fourier coverage for the 12-hole Large Binocular Telescope mask (left) and a hypothetical 12-hole Giant Magellan Telescope mask (right). The top row shows parallactic angle coverage between $-5^\circ$ and $5^\circ$ and the bottom row shows $-65^\circ$ to $65^\circ$.}
\label{fig-gmtuv}
\end{figure}

To compare the current state of the art to future facilities, we simulate observations with the dual-aperture LBT to those for a hypothetical 12-hole mask made for GMT. 
For both masks, we simulate observations with 20 pointings having parallactic angles between $-65^\circ $ and $65^\circ$, comparable to the sky rotation coverage in the December 2014 dataset.
We also simulate observations with poorer sky rotation - two pointings with parallactic angles of $-5^\circ$ and $5^\circ$.
This sky rotation is comparable to the amount observed for the 2012 dual-aperture LBT observations of MWC 349A \citep{2017ApJ...844...22S}.
Figure \ref{fig-gmtuv} shows the Fourier coverage for each facility and sky rotation case.

\begin{figure*}
\epsscale{1.2}
\plotone{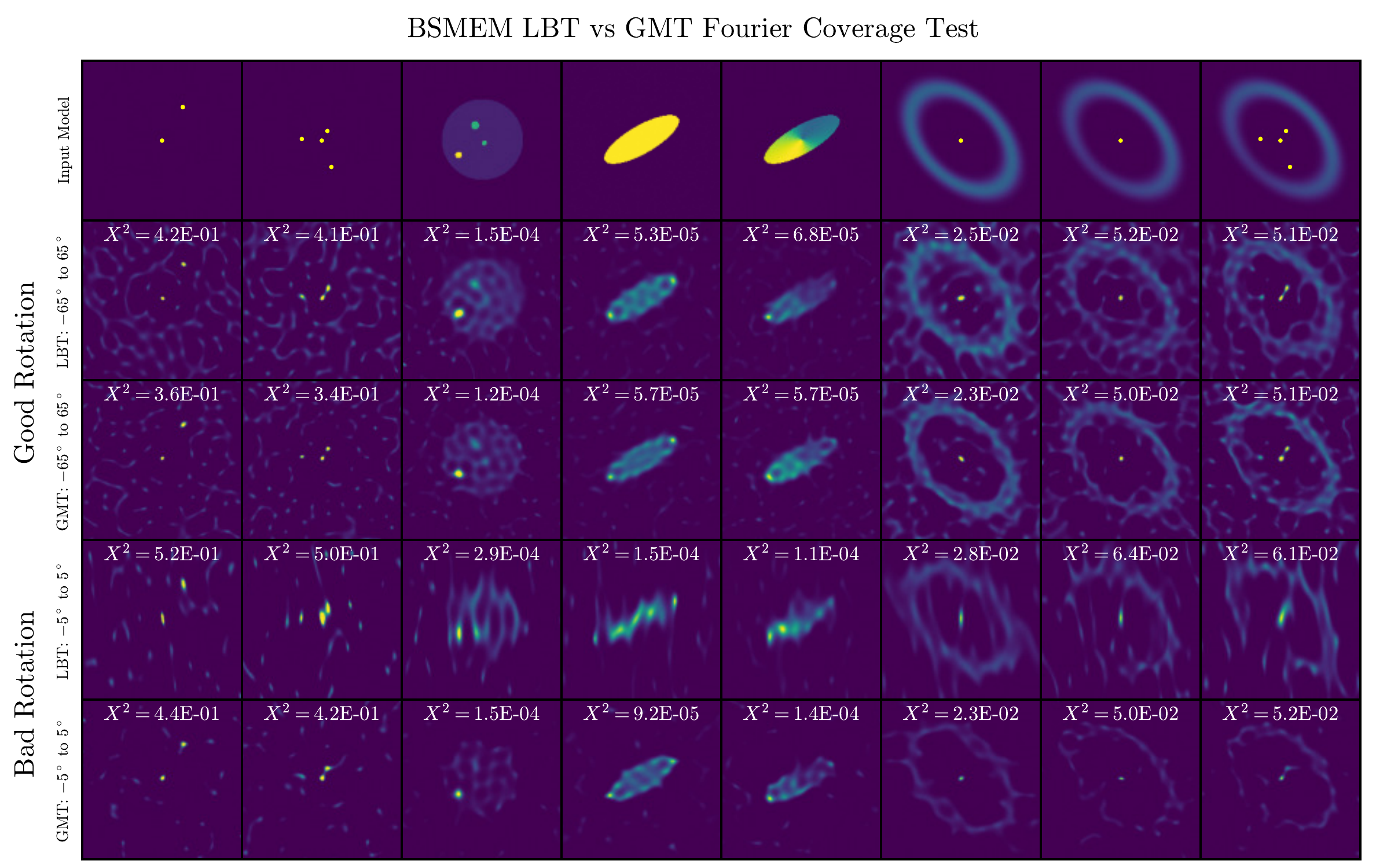}
\caption{Images reconstructed from simulated observations using BSMEM with different masks and sky rotation coverage. The top row shows the input model image. The next two rows show observations with 20 pointings having sky rotation angles between $-65^\circ$ and $65^\circ$ for the LBT and hypothetical GMT masks, respectively. The last two rows show observations with 2 pointings having sky rotation angles of $-5^\circ$ and $5^\circ$ for the LBT and GMT masks. The point sources in Models 1, 2, 6, 7, and 8 in the top row have scaled brightnesses and sizes for ease of viewing. Figure \ref{fig-gmtuv} shows the Fourier coverage of these four datasets.}
\label{fig-gmtims}
\end{figure*}

We reconstruct images for all eight objects in each of these four cases.
Figure \ref{fig-gmtims} shows the results using BSMEM. 
For the good sky rotation case, LBT and GMT produce comparable reconstructed images. 
With poorer sky rotation, the LBT's uneven Fourier coverage leads to poorer reconstructed images.
The lower resolution in one direction smears flux out along the position angle where the beam is wider. 
However, even with smaller sky rotation coverage, none of the reconstructed images are so poor that the input source morphology is not recovered.
Since most datasets will have sky rotation coverage between the ``Good" and ``Bad" cases, Figure \ref{fig-gmtims} shows that the co-phased LBTI can already provide comparable imaging to that possible with GMT.

While LBTI's long baselines can achieve GMT-like resolution now, GMT's symmetric aperture will provide more even Fourier coverage. GMT will thus make detections more efficiently for objects that are unresolved with baselines less than $\sim8$ meters. 
Here, GMT can provide higher quality image reconstructions from observations with less sky rotation (see Figure \ref{fig-gmtims}).
Furthermore, significantly increasing the number of holes in the LBTI mask requires that the holes themselves be much smaller, or else baselines bleed into one another in the Fourier plane.
GMT's larger collecting area would allow for many more holes with the same diameter as the LBTI 12-hole mask.
This would increase efficiency by allowing for shorter exposures and by increasing the amount of recoverable phase information; it would also make GMT more sensitive to fainter sources.

\subsection{Summary: Image Reconstruction Strategies}
The ambiguities described in Sections \ref{sec:degen}, \ref{sec:sims} and in the Appendix show that many factors affect the quality of reconstructed images.
Different algorithms behave differently depending on error bar over- or under-estimation, as well as the choice of priors and baseline weighting schemes.
Furthermore, different sources will reconstruct with different quality depending on their size relative to the array baselines (see Figure \ref{fig-mwc349_sims}), as well as the sky rotation coverage (see Figure \ref{fig-gmtims}).
We thus recommend reconstructing images for simulated data to compare to real observations.
This is useful for understanding what degeneracies and ambiguities exist for particular algorithms and observing strategies.

Particular observation strategies can also be useful for interpreting reconstructed images.
Comparing reconstructed images from multi-wavelength datasets can distinguish between different source morphologies that could lead to similar image reconstructions. 
For example, the position of a companion candidate should remain fixed with wavelength, while quasi-static speckle locations will change with wavelength.
Multi-wavelength observations can also be compared to radiative transfer modeling to differentiate scattered light scenarios from thermal emission.
Furthermore, passive, thermally emitting disk sizes should increase with increasing wavelength, because dust farther from the star will be cooler and emit at longer wavelengths.
Since disks can masquerade as companions for certain array resolutions and sky rotation coverage, this wavelength dependence may be useful for distinguishing between these scenarios.
Multi-epoch datasets are even more powerful for breaking the degeneracies between disks and companions, since static disks cannot cause companion signals with smoothly changing position angles \citep[e.g.][]{2016SPIE.9907E..0DS}.

\section{Conclusions}
We described observational and data reduction strategies for non-redundant masking observations with specific examples for individual datasets.
We showed how image calibrations such as channel bias subtraction, flat fielding, and bad pixel corrections affect the uncalibrated and calibrated observables for LMIRCam.
We recommend checking reduction steps in this way for each dataset since factors such as observing conditions and detector features can change the relative importance of these calibrations.

We explored different closure phase and squared visibility calculation methods.
The closure phases calculated using the unwindowed, ``Monnier" method resulted in lower scatter than all single-pixel calculation methods.
Kernel phase projections that took into account information from the mask alone (the ``Martinache" projection) led to lower scatter and fractional errors than projections that diagonalized the average closure phase covariance matrix.
While windowing did not change the squared visibilities dramatically for observations taken in photometric conditions, datasets with more variable background levels could benefit from windowing.
We described different calibration strategies for the phases and visibilities and made a case for applying simple calibrations rather then optimized ones that have more arbitrary biases.

We presented image reconstruction tests using the BSMEM and SQUEEZE algorithms for eight models representing potential NRM science targets.
Overall, without prior knowledge of the source morphology and with the potential for poorly estimated error bars, BSMEM led to more robust reconstructed images than SQUEEZE.
However, neither algorithm led to such poor reconstructed images that the source morphology could not be identified.
These simulations demonstrated the utility of comparing simulated reconstructions to those for real observations.
We also discussed ways in which multi-wavelength and multi-epoch datasets can be useful for interpreting reconstructed images.

We compared the imaging capabilities of the LBTI 12-hole mask to a hypothetical 12-hole mask designed for GMT.
For good sky rotation coverage and typical noise levels, the dual-aperture LBTI and the GMT produced nearly identical reconstructed images.
The dual aperture LBTI can thus already provide GMT-like NRM imaging.
GMT's larger collecting area and more even Fourier coverage will enable detections at similar angular separations with lower sky rotation and a shorter total integration time.
Given the expected performance of GMT's segment phasing and adaptive optics, NRM remains a promising method for high resolution imaging.

\acknowledgements
This material is based upon work supported by the National Science Foundation under A.A.G. Grant No. 1211329.
This material is based upon work supported by the National Science Foundation under Grant No. 1228509.
This material is based upon work supported by the National Science Foundation Graduate Research Fellowship under Grant No. DGE-1143953. Any opinion, findings, and conclusions or recommendations expressed in this material are those of the authors(s) and do not necessarily reflect the views of the National Science Foundation. 
This research made use of Astropy, a community-developed core Python package for Astronomy \citep{2013A&A...558A..33A}.
This research made use of the open source software packages Scipy \citep{scipy}, Numpy \citep{2011arXiv1102.1523V}, and matplotlib \citep{2007CSE.....9...90H}.

\appendix

\section{Image Reconstructions for Simulated Data}

\begin{figure}
\epsscale{0.5}
\plotone{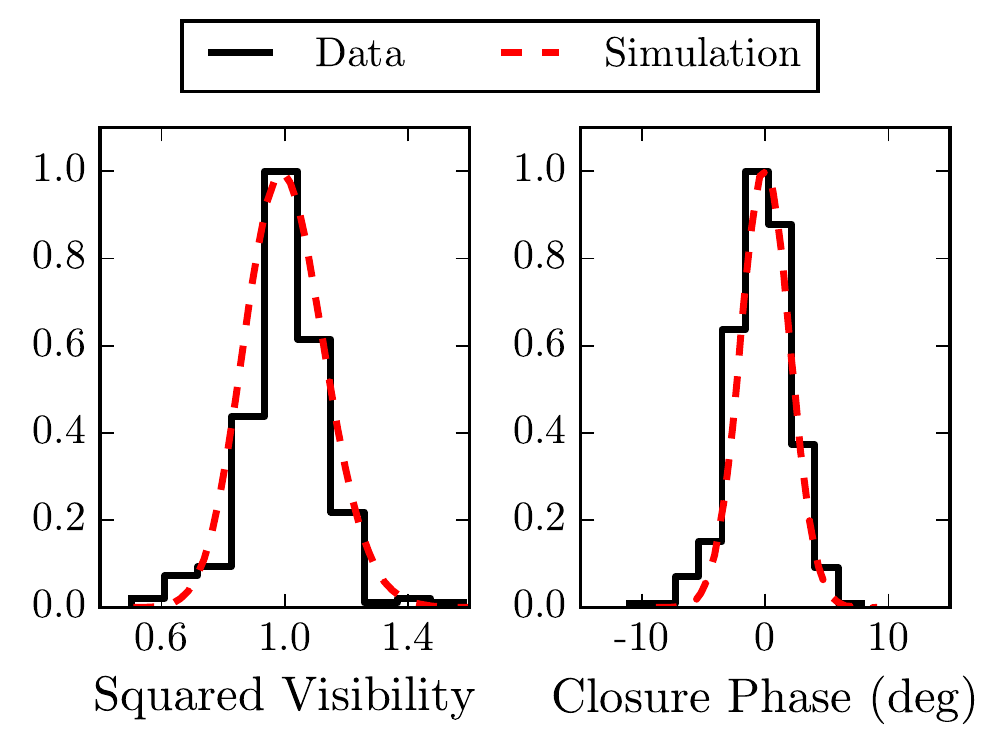}
\caption{Comparison of simulated squared visibilities (left) and closure phases (right) to those observed for the unresolved calibrator stars observed in December 2014. The black histograms show the real observations and the red dashed curves show the simulated data.}
\label{fig-simscat}
\end{figure}

We simulate observations of the following sources to test two image reconstruction algorithms, BSMEM and SQUEEZE, on potential NRM science targets.
\begin{itemize}
\item Model 1: A 1-magnitude contrast binary with a separation of 100 milliarcseconds.
\item Model 2: A multiple system with contrasts ranging from 1 to 5 magnitudes and separations ranging from 30 to 70 milliarcseconds.  Keck surveys of young stars \citep[e.g.][]{2011ApJ...731....8K} detected companions of roughly these angular separations and magnitude ranges (and those in Model 1).
\item Model 3: A uniform circle of diameter 200 mas with spots having contrasts of 2 to 4 magnitudes. NRM is used to image circumstellar disks, which may have non-uniformities such as hot spots. This case is also interesting for long baseline optical interferometry, which is often used to image stellar surfaces.
\item Model 4: A uniform ellipse with a major axis of 200 mas and axis ratio of 0.33.
\item Model 5: A skewed ellipse with a major axis of 200 mas, axis ratio of 0.33, and 50\% skew amplitude. This case and Model 4 are meant to represent the circumstellar disks that may have skew or outflows that have been imaged by NRM studies \citep[e.g.][]{2001ApJ...562..440D,2017ApJ...844...22S}
\item Model 6: A point source plus a uniform ring with a diameter of 300 mas and Gaussian cross section, meant to emulate a transition disk with no detectable companions. This diameter is roughly the angular diameter of transition disk clearings for the most nearby potential targets, corresponding to a hole diameter of 84 AU at 140 pc \citep[e.g.][]{2011ApJ...732...42A}. 
\item Model 7: A point source plus a skewed Gaussian ring with a diameter of 300 mas and 70\% skew amplitude. Skew has been observed in NRM observations of transition disks \citep[e.g.][]{2011A&A...528L...7H,2015ApJ...801...85S,2015MNRAS.450L...1C}.
\item Model 8: The multiple system in Model 2 surrounded by the skewed Gaussian ring in Model 7. This is meant to imitate companion searches in skewed transition disks. Companions with separations and contrasts similar to those in Model 2 have been found in NRM observations of transition and circumbinary disks \citep[e.g.][]{2008ApJ...678L..59I,2012ApJ...745....5K,2015Natur.527..342S}. 
\end{itemize}
All datasets are generated using the 12-hole mask installed in LBTI/LMIRCam in dual-aperture mode.
To create data with realistic amounts of correlated noise, we add Gaussian noise to the complex visibility amplitudes $\left( A \right)$ and phases $\left( \phi \right)$ before calculating the squared visibilities and closure phases. 
As shown in Figure \ref{fig-simscat}, amplitude and phase scatters of 0.065 and $1.15^\circ$ match the observed scatter in the December 2014 dataset.
The propagated uncertainties for the closure phases and squared visibilities are then
\begin{equation} 
\sigma_{CP} = \sqrt{3}~\sigma_\phi
\label{eq-cperr}
\end{equation}
and
\begin{equation}
\sigma_{V^2} = \sqrt{2A}~\sigma_A.
\label{eq-v2err}
\end{equation}
Unless specified otherwise, we use these noise levels and roughly the same sky rotation coverage as in the December 2014 dataset - 20 pointings with parallactic angles between $-65^\circ$ and $65^\circ$.

We reconstruct images using the BSMEM \citep{1994IAUS..158...91B} and SQUEEZE \citep{2010SPIE.7734E..2IB} algorithms.
The total image field of view is always set to 780 mas, corresponding to the resolution of the shortest baseline in the mask, and the pixel scale is always set to 2 mas.
To calculate residuals, we first align the output image with the input model via cross-correlation and then sum the square of the images' difference. 
We perform BSMEM reconstructions in ``Classic Bayesian" mode, running 200 iterations for every input dataset.
For SQUEEZE, we run 4 MCMC chains in parallel tempering mode for 1000 iterations.
Since the point of these imaging experiments is to test the algorithms with no knowledge of the input source, we do not optimize any regularization or noise scalings when reconstructing images. 
The results of all the reconstructions for both algorithms are discussed here, with the images shown in Figures \ref{fig-priorbs} to \ref{fig-cweightss}. 

\subsection{Initial Images}

\begin{figure*}
\epsscale{1.2}
\plotone{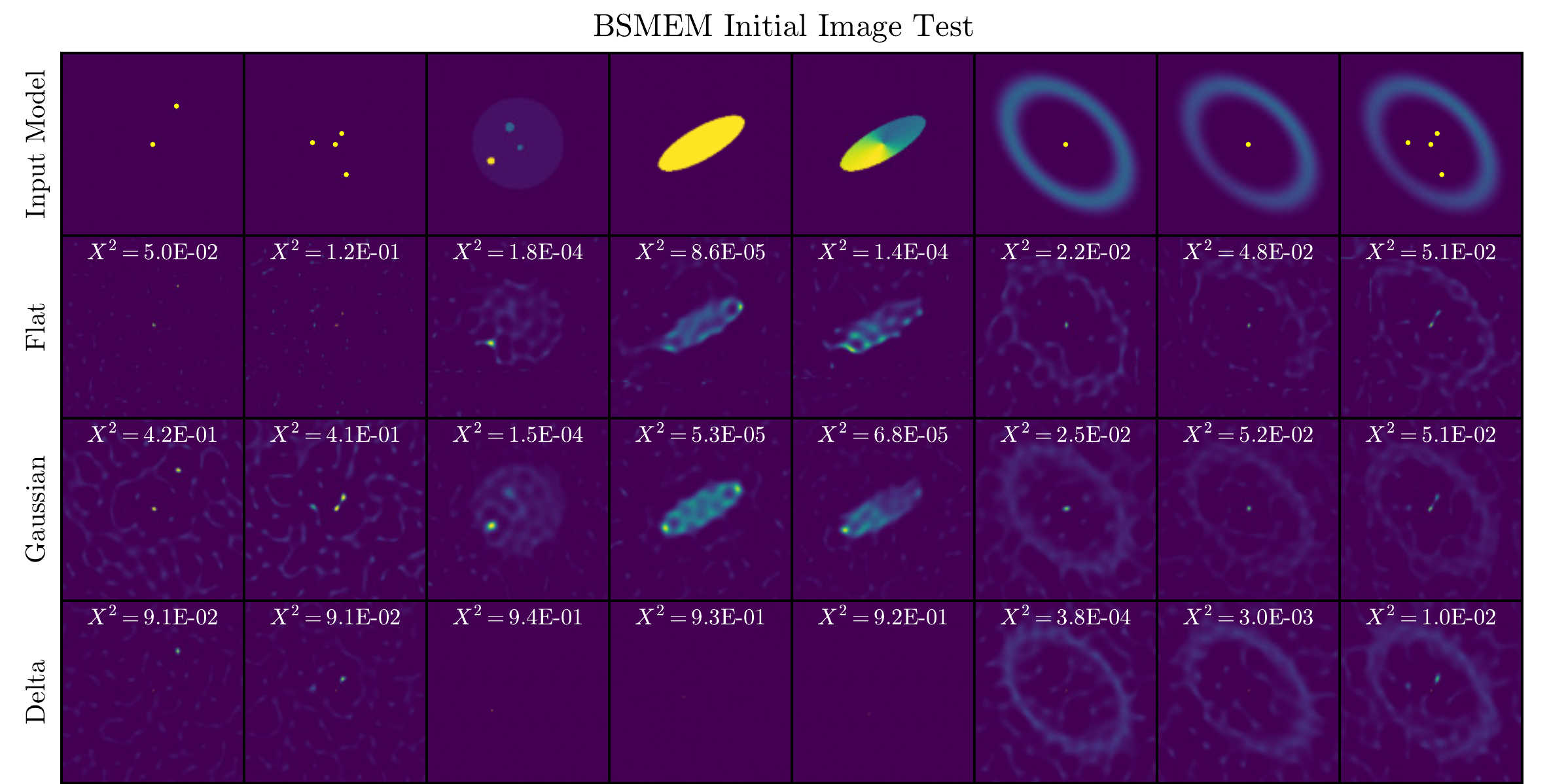}
\caption{Images reconstructed from simulated observations using BSMEM with different initial images. The top row shows the input model image, and the next three rows show reconstructions with flat, Gaussian, and delta function initial images, respectively. The point sources in Models 1, 2, 6, 7, and 8 in the top row have scaled brightnesses and sizes for ease of viewing.}
\label{fig-priorbs}
\end{figure*}

We test the effect of using different starting images (flat, Gaussian, and $\delta$ functions) in both BSMEM and SQUEEZE for each of the eight sources.
The Gaussian initial images have full-width-half-maxima of 200 mas.
Figure \ref{fig-priorbs} shows the results for BSMEM.
Here, starting with a flat initial image or a centered Gaussian provides the best reconstruction with no knowledge of the true source morphology.
The binary and multiple reconstructions have lower residuals with a delta function initial image.
However, for extended sources, BSMEM cannot converge to a realistic image in classic Bayesian mode with a delta function prior.
The flat initial image results also have vertical or horizontal striping in some cases.
This is because BSMEM does not automatically recenter when reconstructing images non-interactively. 
Recentering by hand during the reconstruction would eliminate these artifacts. 

\begin{figure*}
\epsscale{1.2}
\plotone{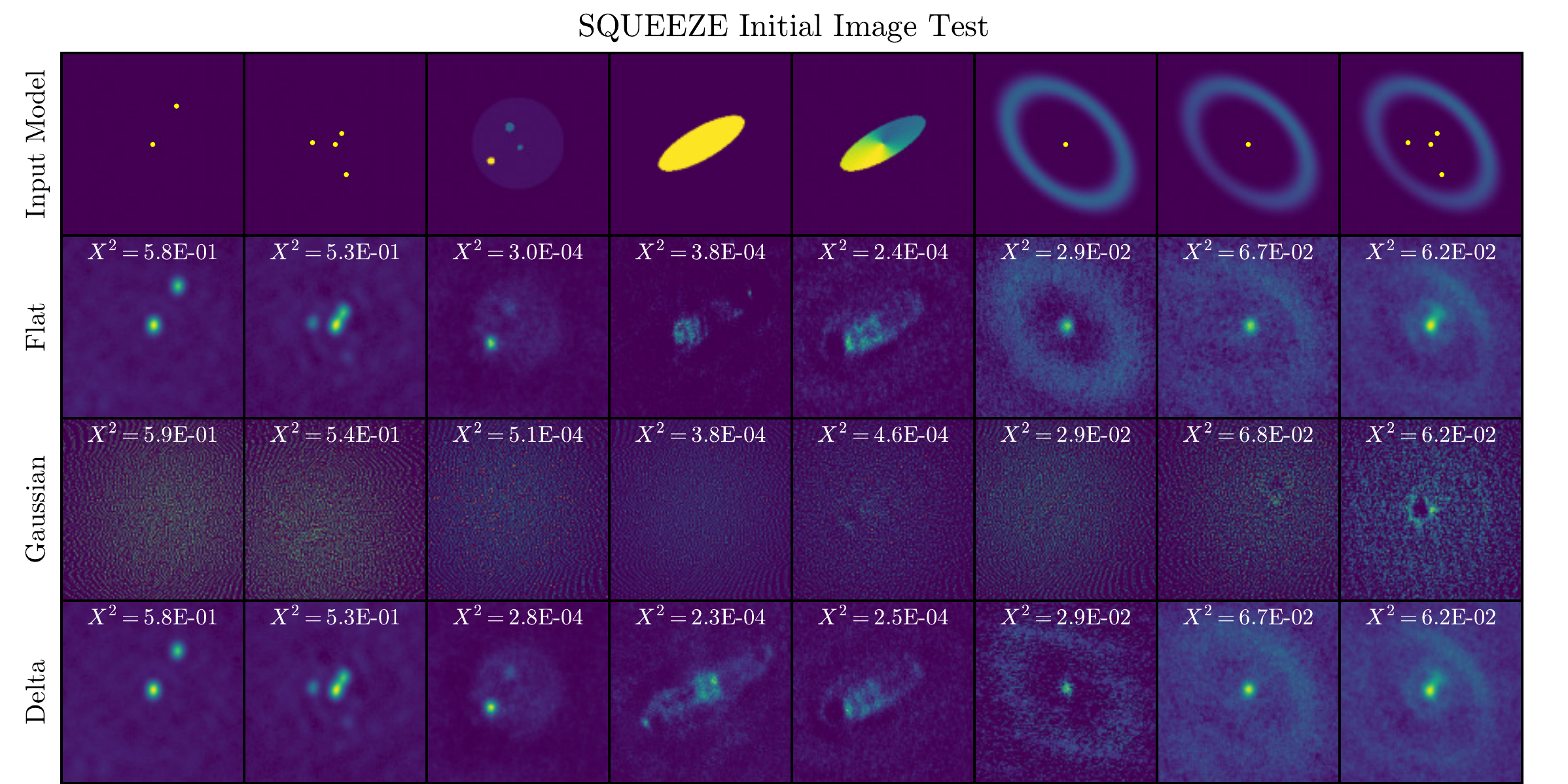}
\caption{Images reconstructed from simulated observations using SQUEEZE with different initial images (rows). The top row shows the input model image, and the next three rows show reconstructions with flat, Gaussian, and delta function initial images, respectively. The point sources in Models 1, 2, 6, 7, and 8 in the top row have scaled brightnesses and sizes for ease of viewing.}
\label{fig-priors}
\end{figure*}

Figure \ref{fig-priors} shows the same tests using SQUEEZE.
Here, only the flat and $\delta$ function initial images result in reconstructions that match the input images.
With 1000 steps, SQUEEZE does not reproduce the input model starting from a Gaussian initial image.
Comparing the reconstructions of models that include compact point sources (i.e. Models 1, 2, and 8) between BSMEM and SQUEEZE shows that SQUEEZE is more likely to blend point sources together.
As a result, BSMEM more reliably reproduces the point sources within the ring in Model 8.
Conversely, comparing Models 3, 4, and 5 between the two algorithms shows that SQUEEZE is less prone to over-resolve extended emission than BSMEM.

\subsection{Total $\chi^2$ Scaling}
We scale the total $\chi^2$ of each dataset to explore the dependence of each algorithm on correct error bar estimation.
We multiply all the closure phase and squared visibility error bars by constant values of 0.25, 0.5, 1.0, 2.0, and 4.0, without changing the actual noise added to the model.
For each algorithm we use the initial image that produced the best results in the previous section - a large Gaussian for BSMEM, and a flat image for SQUEEZE.

\begin{figure*}
\epsscale{1.2}
\plotone{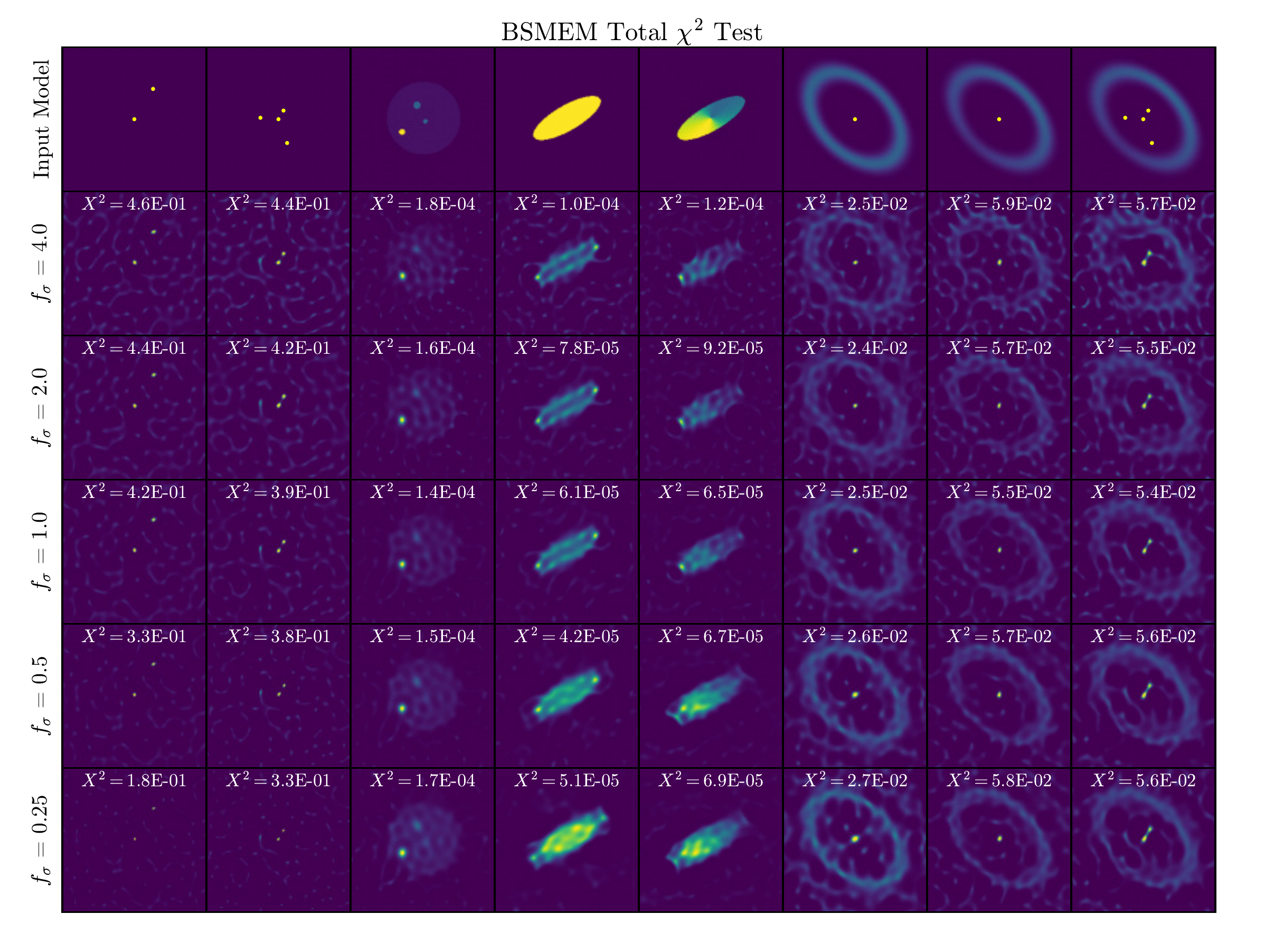}
\caption{Images reconstructed from simulated observations using BSMEM with different error bar scalings. The top row shows the input model image, and the following rows have all closure phase and squared visibility errors scaled by factors of 4.0, 2.0, 1.0, 0.5, and 0.25 from top to bottom. The point sources in Models 1, 2, 6, 7, and 8 in the top row have scaled brightnesses and sizes for ease of viewing.}
\label{fig-chi2bs}
\end{figure*}

\begin{figure*}
\epsscale{1.2}
\plotone{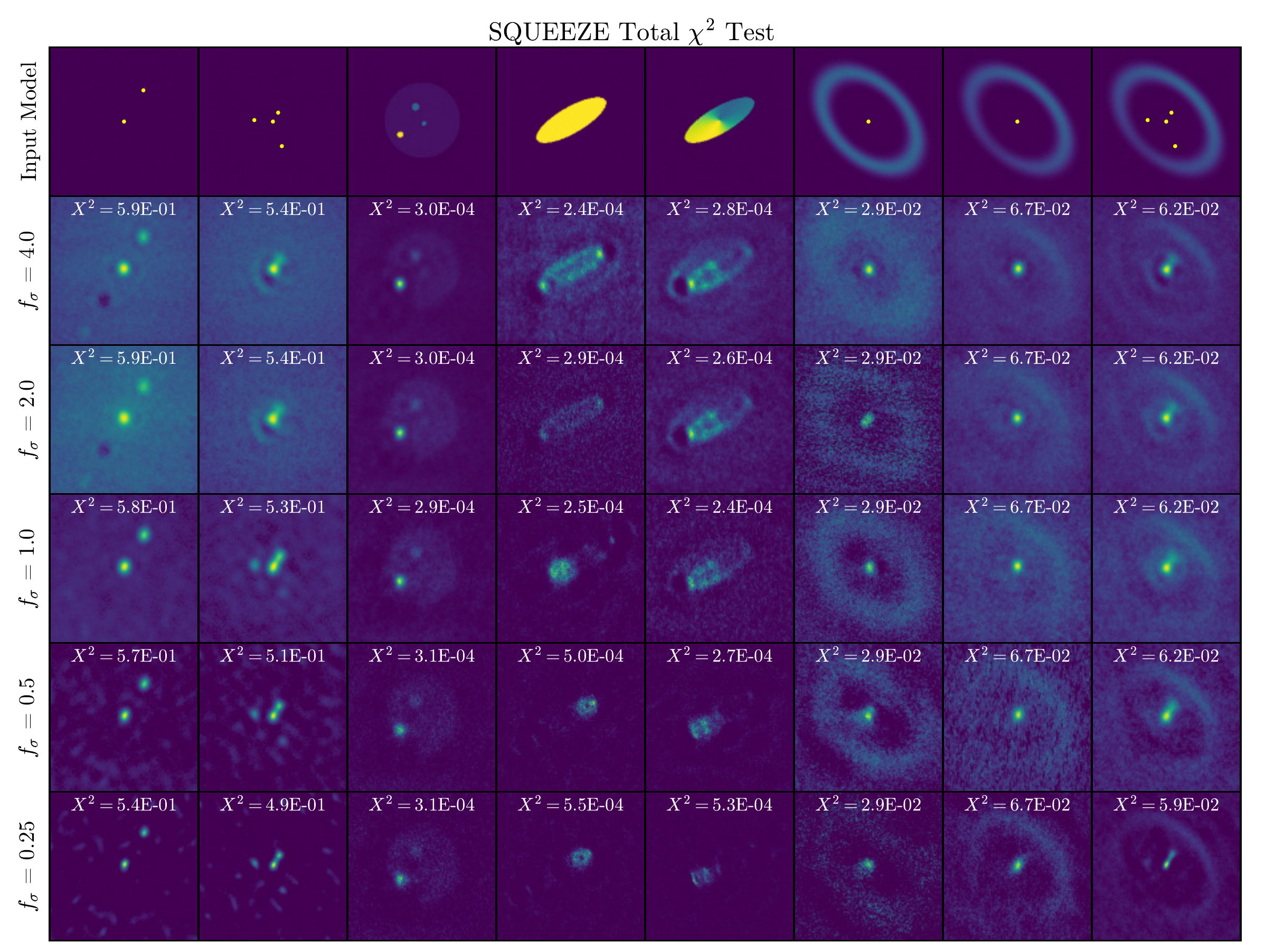}
\caption{Images reconstructed from simulated observations using SQUEEZE with different error bar scalings. The top row shows the input model image, and the following rows have all closure phase and squared visibility errors scaled by factors of 4.0, 2.0, 1.0, 0.5, and 0.25 from top to bottom. The point sources in Models 1, 2, 6, 7, and 8 in the top row have scaled brightnesses and sizes for ease of viewing.}
\label{fig-chi2s}
\end{figure*}

Figures \ref{fig-chi2bs} and \ref{fig-chi2s} show the results for BSMEM and SQUEEZE, respectively. 
As the error bar scaling decreases, both algorithms produce tighter reconstructions of the point sources in Models 1, 2, and 8. 
With over-estimated error bars, SQUEEZE's blurring of close-separation point sources becomes more pronounced (see Model 8 in Figure \ref{fig-chi2s}). 
In this regime, SQUEEZE also puts regions of very low flux directly opposite asymmetric emission (see Model 1, top two panels in Figure \ref{fig-chi2s}).
While SQUEEZE blurs extended emission (e.g. Models 3, 4 and 5) more with over-estimated errors, BSMEM creates blurrier reconstructions of extended emission when the errors are under-estimated.

Neither algorithm causes huge qualitative mismatches between the inputs and reconstructed images for any of the error scalings.
This suggests that poorly estimated errors will not degrade reconstructed images to the point that the source morphology could not be recovered.
However, for SQUEEZE especially, poorly estimated error bars may result in non-detections of close-in companions.
Comparing Figures \ref{fig-chi2bs} and \ref{fig-chi2s} shows that BSMEM produces more reliable reconstructions of most models with under- or over-estimated errors.
 
\subsection{Weighting By Baseline Length}

To test the effects of different ($u,v$) weighting schemes, we reconstruct images using closure phase and squared visibility errors that depend on baseline length.
We assign errors in the following way:
\begin{equation}
\sigma = B^p~ \sigma_{med},
\end{equation}
where B is the baseline length for squared visibilities and the mean baseline length for closure phases, and $\sigma_{med}$ is the median observed error bar.
Here, $p = 0$ corresponds to uniform weighting, $p < 0$ upweights long baselines, and $p > 0$ upweights short baselines.
To separate the relative weighting effects from a varying total $\chi^2$, we scale the errors so that the total $\chi^2$ of the initial image (Gaussian for BSMEM and flat for SQUEEZE) is constant. 
We reconstruct images for p values of -2, -1, 0, 1, 2.

\begin{figure*}
\epsscale{1.2}
\plotone{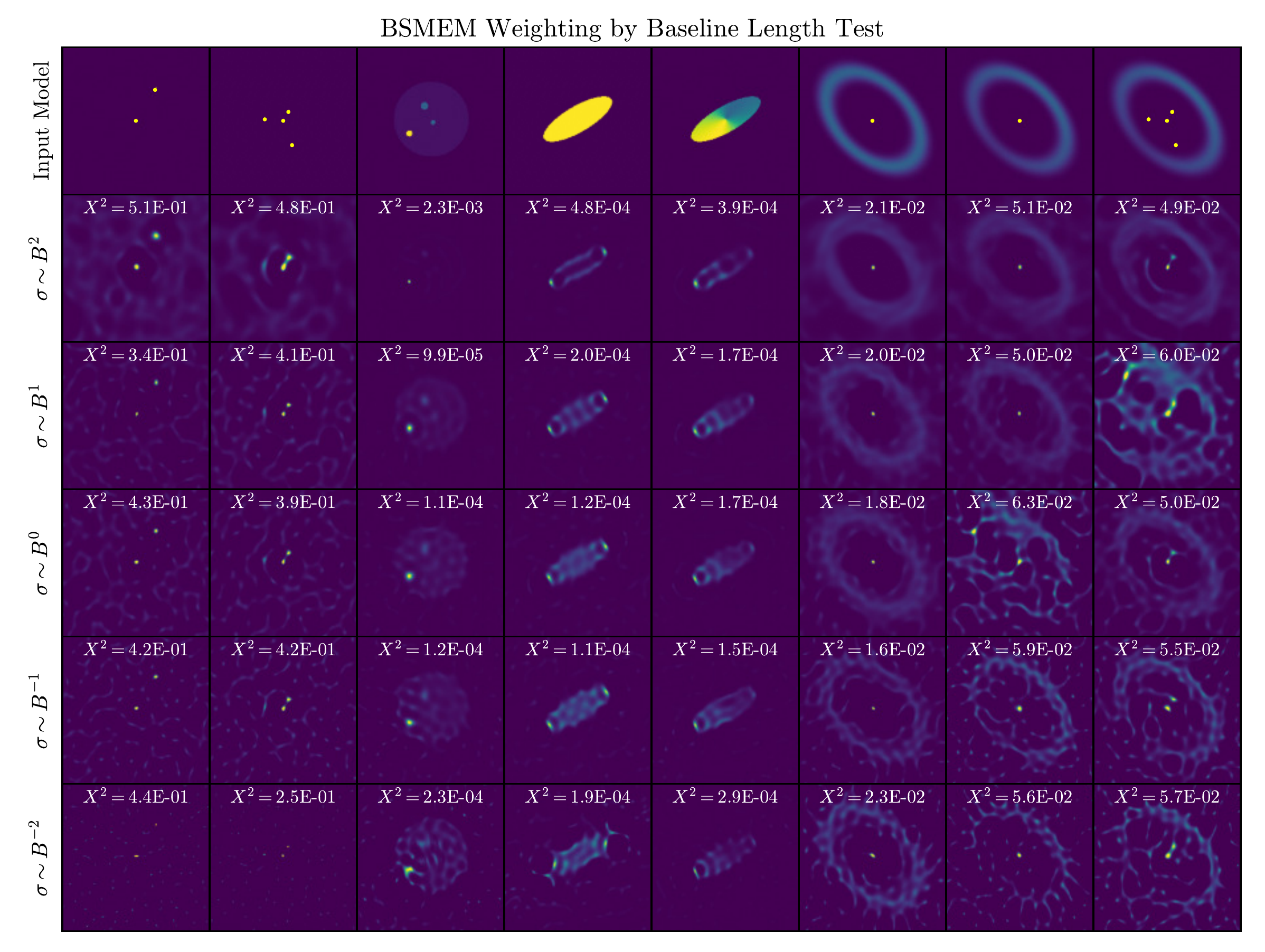}
\caption{Images reconstructed from simulated observations using BSMEM with different baseline weightings. The top row shows the input model image. The next rows show reconstructions for datasets whose assigned errors depend on baseline length raised to powers of 2, 1, 0, -1, and -2 from top to bottom. The upper rows weight short baselines more heavily, while the lower rows weight longer baselines more heavily. The point sources in Models 1, 2, 6, 7, and 8 in the top row have scaled brightnesses and sizes for ease of viewing.}
\label{fig-bweightsbs}
\end{figure*}

\begin{figure*}
\epsscale{1.2}
\plotone{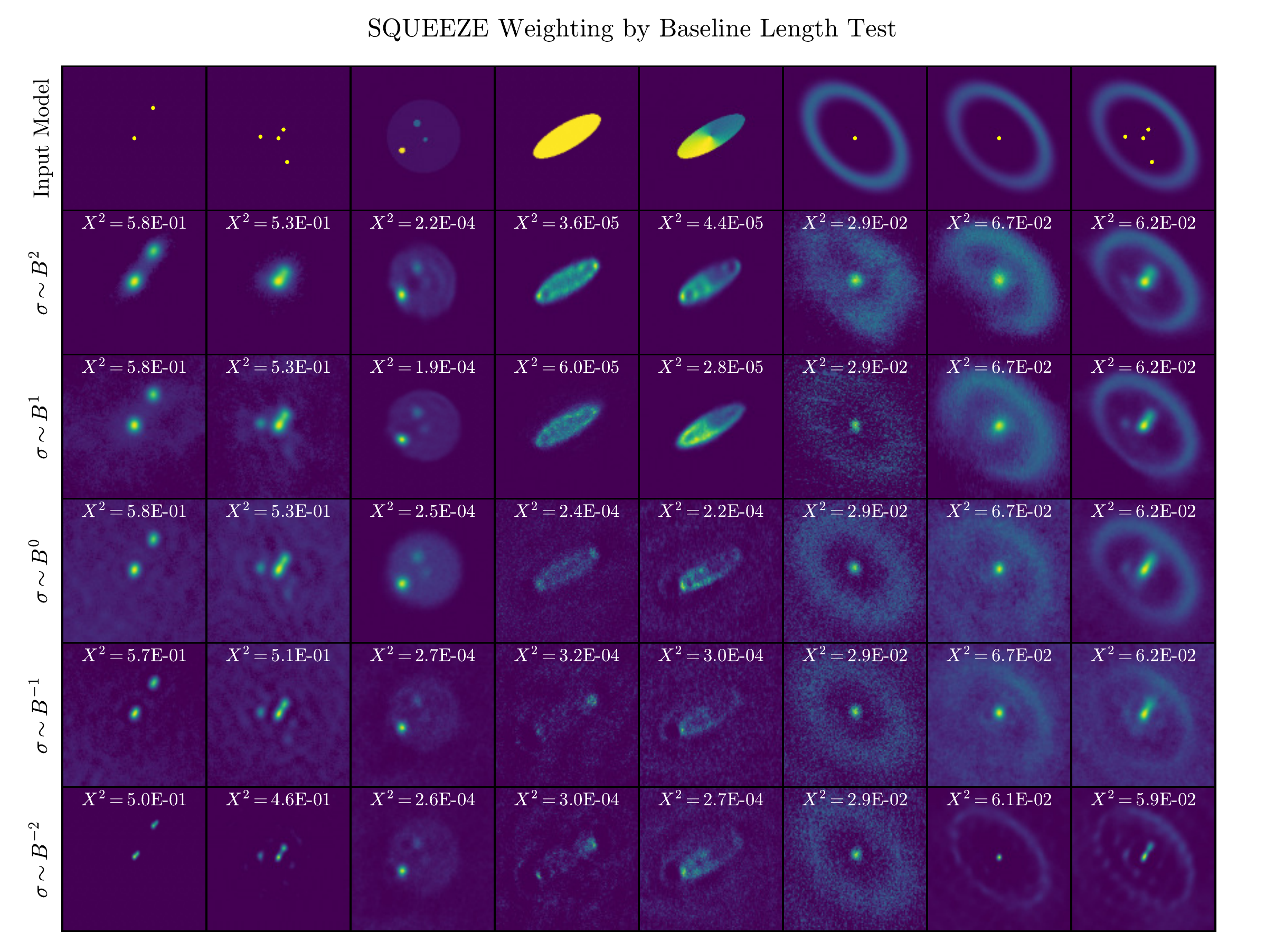}
\caption{Images reconstructed from simulated observations using SQUEEZE with different baseline weightings. The top row shows the input model image. The next rows show reconstructions for datasets whose assigned errors depend on baseline length raised to powers of 2, 1, 0, -1, and -2 from top to bottom. The upper rows weight short baselines more heavily, while the lower rows weight longer baselines more heavily. The point sources in Models 1, 2, 6, 7, and 8 in the top row have scaled brightnesses and sizes for ease of viewing.}
\label{fig-bweightss}
\end{figure*}

Figure \ref{fig-bweightsbs} shows the results for BSMEM, and Figure \ref{fig-bweightss} shows the results for SQUEEZE.
Both algorithms generally produce better reconstructions of extended emission when the short baselines are upweighted.
SQUEEZE's reconstructions of Models 4 and 5 show a more marked improvement with upweighted short baselines than BSMEM's.
For both algorithms, upweighting the long baselines improves the recovery of compact sources, especially when they are alongside an extended component like in Model 8 (skewed ring + multiple system).
However, both SQUEEZE and BSMEM tend to over-resolve extended components when the long baselines are upweighted (see bottom rows of Figures \ref{fig-bweightsbs} and \ref{fig-bweightss}).
As in the total $\chi^2$ tests, no weighting scheme creates large qualitative differences between the inputs and reconstructions.

\subsection{Closure Phase and Squared Visibility Weighting}
We test the effect of reconstructing images using different relative closure phase and squared visibility weights.
We scale all closure phase errors by a constant factor $f_{cp}$, while leaving the squared visibility errors alone ($f_{V^2} = 1.0$).
We then rescale $f_{cp}$ and $f_{V^2}$ by the same constant to keep the $\chi^2$ of the prior image for each algorithm fixed.
We use error scaling ratios ($f_{cp} / f_{V^2}$) of 4.0, 2.0, 1.0, 0.5, and 0.25, where larger values downweight closure phases and smaller values upweight closure phases relative to squared visibilities.

\begin{figure*}
\epsscale{1.2}
\plotone{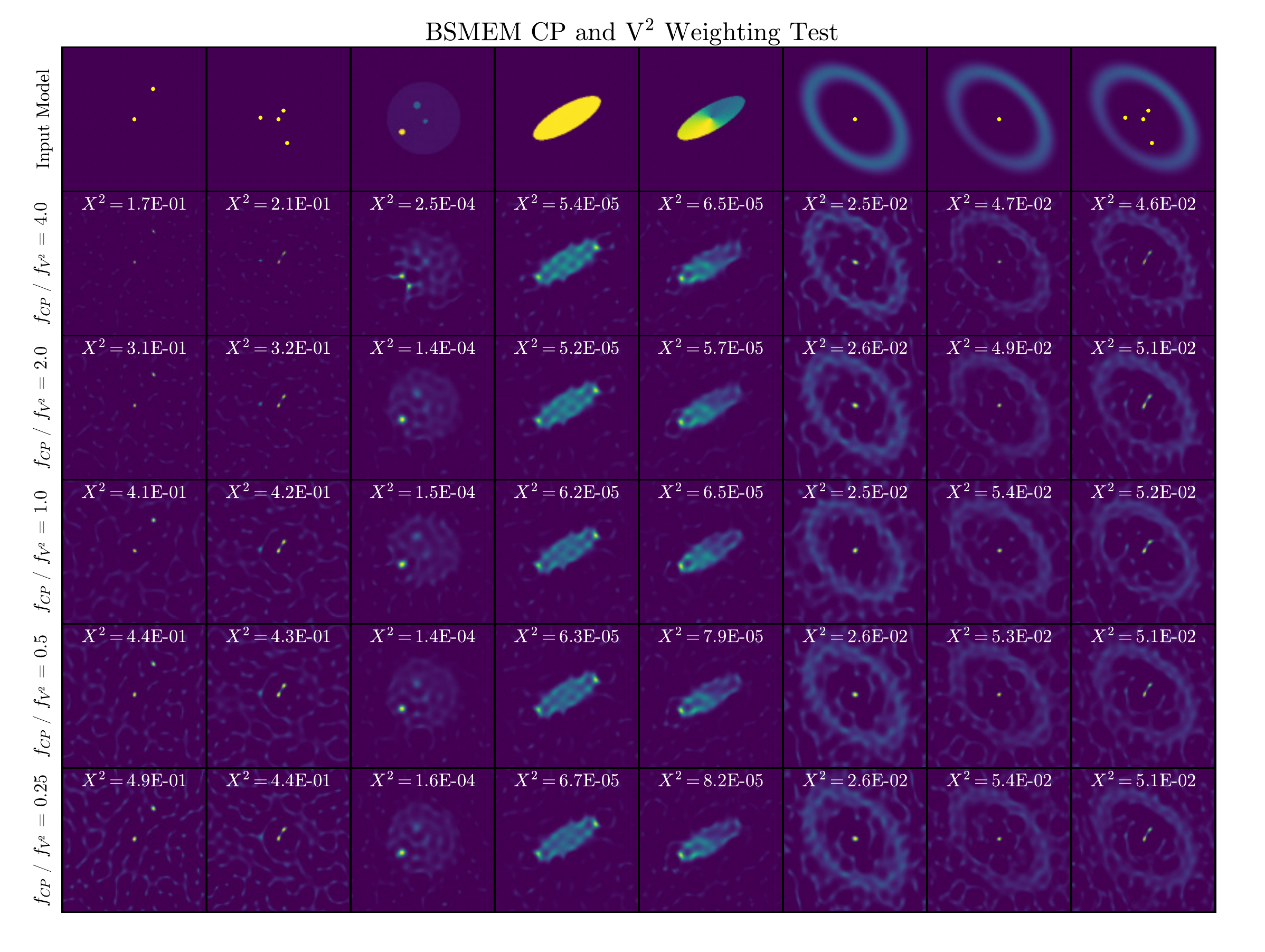}
\caption{Images reconstructed from simulated observations using BSMEM with different relative closure phase and squared visibility weightings. The top row shows the input model image, and the next rows show reconstructions where the closure phase errors have been scaled by a factor of 4.0, 2.0, 1.0, 0.5, and 0.25 top to bottom, relative to the squared visibilities. The point sources in Models 1, 2, 6, 7, and 8 in the top row have scaled brightnesses and sizes for ease of viewing.}
\label{fig-cweightsbs}
\end{figure*}

\begin{figure*}
\epsscale{1.2}
\plotone{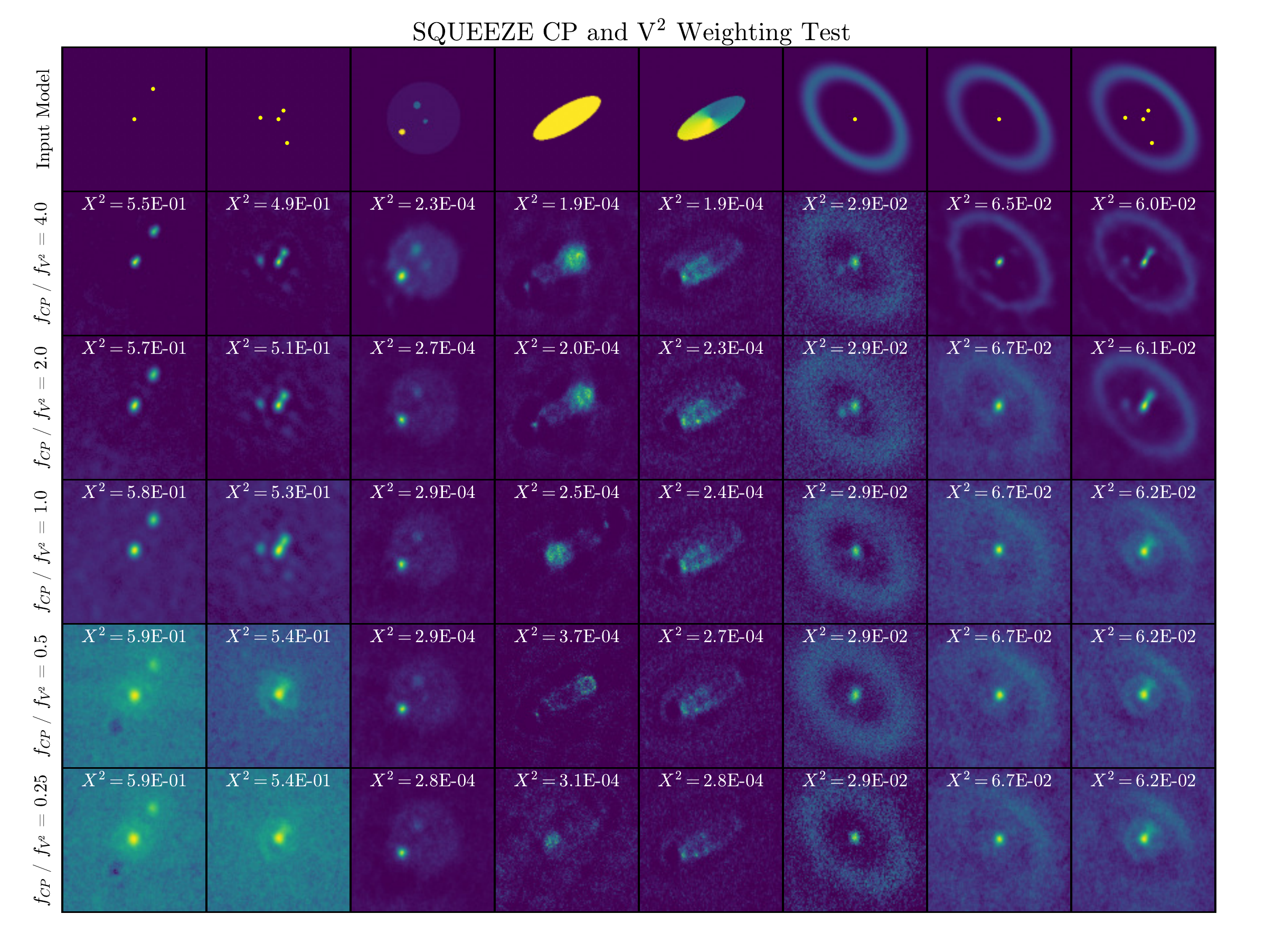}
\caption{Images reconstructed from simulated observations using BSMEM with different relative closure phase and squared visibility weightings. The top row shows the input model image, and the next rows show reconstructions where the closure phase errors have been scaled by a factor of 4.0, 2.0, 1.0, 0.5, and 0.25 top to bottom, relative to the squared visibilities. The point sources in Models 1, 2, 6, 7, and 8 in the top row have scaled brightnesses and sizes for ease of viewing.}
\label{fig-cweightss}
\end{figure*}

Figures \ref{fig-cweightsbs} and \ref{fig-cweightss} show the results for BSMEM and SQUEEZE, respectively.
BSMEM shows little to no difference in reconstruction quality across all weighting schemes.
The exception to this is the $f_{cp} / f_{V^2} = 4$ reconstruction of Model 3, which has a poorer reconstruction of the spots on the uniform disk.

SQUEEZE produces more reliable reconstructions of Models 1, 2, 7, and 8 when the closure phases are downweighted relative to the squared visibilities. 
It does not recover the compact components in these models when the squared visibilities are downweighted.
However, the reconstructions of Model 6 (uniform ring + delta function) are worse with downweighted closure phases.
Here SQUEEZE allows compact emission to appear in two locations within the ring.
This could result from averaging subsequent images in the MCMC chains that were not centered consistently.
Model 4 appears particularly asymmetric for large closure phase errors.
Increasing the power law exponent of the chain temperature distribution, and thus allowing SQUEEZE to explore larger parts of parameter space more quickly, makes the reconstructions more symmetric.
This suggests that it takes longer for SQUEEZE to converge to the true source morphology when there is not much information in the closure phases.
Overall, the relative closure phase and squared visibility weights change SQUEEZE's performance more dramatically than BSMEM's.

\clearpage

\bibliography{/Users/stephaniesallum/Papers/PaperWritingTools/Latex/references}
\end{document}